\documentclass[10pt,onecolumn,letterpaper,twoside]{IEEEtran}
\usepackage{graphicx}				
\usepackage{amssymb,amsmath}
\usepackage{algorithm}
\usepackage{color}

\newtheorem{Assumption}{Assumption}
\newtheorem{Def}{Definition}
\newtheorem{Thm}{Theorem}
\newtheorem{Fact}{Fact}
\newtheorem{Lem}{Lemma}
\newtheorem{Cor}{Corollary}

\newcommand*{\tranT}{{\mkern-1.5mu\mathsf{T}}}
\newcommand*{\tran}{^{\mkern-1.5mu\mathsf{T}}}
\newcommand*{\dom}{\mbox{dom}} 

\title{A New Backpressure Algorithm for Joint Rate Control and Routing with Vanishing Utility Optimality Gaps and Finite Queue Lengths}

\author{Hao Yu and Michael J. Neely \\Department of Electrical Engineering \\
University of Southern California}

\begin{document}
\maketitle

\begin{abstract}
The backpressure algorithm has been widely used as a distributed solution to the problem of joint rate control and routing in multi-hop data networks.  By controlling a parameter $V$ in the algorithm, the backpressure algorithm can achieve an arbitrarily small utility optimality gap. However, this in turn brings in a large queue length at each node and hence causes large network delay. This phenomenon is known as the fundamental utility-delay tradeoff.  The best known utility-delay tradeoff for general networks is $[O(1/V), O(V)]$ and is attained by a backpressure algorithm based on a drift-plus-penalty technique.  This may suggest that to achieve an arbitrarily small utility optimality gap, the existing backpressure algorithms necessarily yield an arbitrarily large queue length.  However, this paper proposes a new backpressure algorithm that has a vanishing utility optimality gap, so utility converges to exact optimality as the algorithm keeps running, while queue lengths are bounded throughout by a finite  constant.  The technique uses backpressure and drift concepts with a new method for convex programming. 
\end{abstract}


\section{Introduction}

In multi-hop data networks, the problem of joint rate control and routing is to accept data into the network  to maximize certain utilities and to make routing decisions at each node such that all accepted data are delivered to intended destinations without overflowing any queue in intermediate nodes.  The original backpressure algorithm proposed in the seminal work \cite{Tassiulas92TAC} by Tassiulas and Ephremides addresses this problem by assuming that incoming data are given and are inside the network stability region and develops a routing strategy to deliver all incoming data without overflowing any queue.  In the context of \cite{Tassiulas92TAC}, there is essentially no utility maximization consideration in the network. The backpressure algorithm is further extended by a drift-plus-penalty technique to deal with data network with both utility maximization and queue stability considerations \cite{PhD_Thesis_Neely,GeorgiadisNeelyTassiulas06, book_Neely10}.  Alternative extensions for both utility maximization and queue stabilization are developed in \cite{Eryilmaz06JSAC,Stolyar05,LinShroff04CDC,Shroff06TWC}. The above extended backpressure algorithms have different dynamics and/or may yield different utility-delay tradeoff results. However, all of them rely on ``backpressure" quantities, which are the differential backlogs between neighboring nodes.

It has been observed in \cite{Neely05JSAC,Eryilmaz06JSAC,LinShroff04CDC,LiuShroff15TON} that the drift-plus-penalty and other alternative algorithms can be interpreted as first order Lagrangian dual type methods for constrained optimization. In addition, these backpressure algorithms follow certain fundamental utility-delay tradeoffs.  For instance, the primal-dual type backpressure algorithm in \cite{Eryilmaz06JSAC} achieves an $O(1/V)$ utility optimality  gap  with an $O(V^2)$ queue length, where $V$ is an algorithm parameter. By controlling parameter $V$, a small utility optimality gap is available only at the cost of a large queue length. The drift-plus-penalty backpressure algorithm \cite{book_Neely10}, which has the best utility-delay tradeoff among all existing first order Lagrangian dual type methods for general networks, can only achieve an $O(1/V)$ utility optimality gap with an $O(V)$ queue length.  Under certain restrictive assumptions over the network, a better $[O(1/V), O(\log(V))]$ tradeoff is achieved  via an exponential Lyapunov function  in  \cite{Neely06JSAC}, and an  $[O(1/V), O(\log^2(V))]$ tradeoff is achieved  via a LIFO-backpressure algorithm in  \cite{HuangNeely13TAC}. The existing utility-delay tradeoff results seem to suggest that a large queueing delay is unavoidable if a small utility optimality gap is demanded.

Recently, there have been many attempts in obtaining new variations of backpressure algorithms by applying Newton's method to the Lagrangian dual function. In the recent work \cite{LiuShroff15TON}, the authors develop a Newton's method for joint rate control and routing. However, the utility-delay tradeoff in \cite{LiuShroff15TON} is still $[O(1/V),O(V^2)]$; and the algorithm requires a centralized projection step (although Newton directions can be approximated in a distributed manner). Work \cite{Wei13TAC-1} considers a network flow control problem where the path of each flow is given (and hence there is no routing part in the problem), and proposes a decentralized Newton based algorithm 
for rate control. Work \cite{Zargham13Globecom} considers network routing without an end-to-end utility and only shows the stability of the proposed Newton based backpressure algorithm. All of the above Netwon's method based algorithms rely on distributed approximations for the inverse of Hessians, whose computations still require certain coordinations for the local information updates and propagations and do not scale well with the network size. In contrast, the first order Lagrangian dual type methods do not need global network topology information.  Rather, each node only needs the queue length information of its neighbors.

This paper proposes a new first order Lagrangian dual type backpressure algorithm that is as simple as the existing algorithms in \cite{book_Neely10,Eryilmaz06JSAC,LinShroff04CDC}  but has a better utility-delay tradeoff.  The new backpressue algorithm achieves a vanishing utility optimality gap that decays like $O(1/t)$, where $t$ is the number of iterations.  It also guarantees that the queue length at each node is always bounded by a fixed constant of the same order as the optimal Lagrange multiplier of the network optimization problem.  This improves 
on the utility-delay tradeoffs of prior work.  In particular, it  improves the $[O(1/V), O(V^2)]$ utility-delay 
tradeoff in \cite{Eryilmaz06JSAC} and  the $[O(1/V), O(V)]$ utility-delay tradeoff of the drift-plus-penalty algorithm in \cite{book_Neely10}, both of which yield an unbounded queue length to have a vanishing utility optimality gap.   The new backpressure algorithm differs from existing  first order backpressure algorithms in the following aspects:
\begin{enumerate}
\item The ``backpressure'' quantities in this paper are with respect to  newly introduced weights. These  
are different from queues used in other backpressure algorithms,  but can still be locally tracked and updated.

\item The rate control and routing decision rule involves a quadratic term that is similar to a term used in  proximal algorithms \cite{Parikh13ProximalAlgorithm}. 
\end{enumerate}

Note that the benefit of introducing a quadratic term in network optimization has been observed in \cite{LinShroff06TAC}. 
Work \cite{LinShroff06TAC} considers a network utility maximization problem with given routing paths 
that is a special case of the problem treated in this paper. The algorithm of \cite{LinShroff06TAC} considers  
a fixed set of predetermined paths for each session  and does not  
scale well when treating all (typically exponentially many) possible paths of a general network.  
  The algorithm proposed  in \cite{LinShroff06TAC} is not a backpressure type and hence is fundamentally different from ours. For example, the  algorithm in \cite{LinShroff06TAC} needs to update the primal variables (source session rates for each path) at least twice per iteration, while our algorithm only updates the primal variables (source session rates and link session rates) once per iteration.  The prior work \cite{LinShroff06TAC} shows that the utility optimality gap is asymptotically zero without analyzing the decay rate,  while this paper shows the utility optimality gap decays like $O(1/t)$.

\section{System Model and Problem Formulation}
Consider a slotted data network with normalized time slots $t\in\{0,1,2,\ldots\}$. This network is represented by a graph $\mathcal{G}=(\mathcal{N}, \mathcal{L})$, where $\mathcal{N}$ is the set of nodes and $\mathcal{L}\subseteq \mathcal{N}\times \mathcal{N}$ is the set of directed links. Let $\vert \mathcal{N}\vert =N$ and $\vert \mathcal{L}\vert = L$. This network is shared by $F$ end-to-end sessions denoted by a set $\mathcal{F}$. For each end-to-end session $f\in \mathcal{F}$, the source node $\text{Src}(f)$ and destination node $\text{Dst}(f)$ are given but the routes are not specified.  Each session $f$ has a continuous and concave utility function  $U_f(x_f)$ that represents the ``satisfaction'' received by accepting $x_f$ amount of data for session $f$ into the network at each slot.  Unlike \cite{Eryilmaz06JSAC,LiuShroff15TON} where $U_f(\cdot)$ is assumed to be differentiable and strongly concave, this paper considers general concave utility functions $U_f(\cdot)$, including those that are neither differentiable nor strongly concave.   Formally, each utility function $U_f$ is defined over an interval $\dom(U_f)$, called the \emph{domain} of the function.  It is assumed throughout that either $\dom(U_f) = [0, \infty)$ or $\dom(U_f) = (0,\infty)$, the 
latter being important for \emph{proportionally fair utilities} \cite{Kelly98JORS} $U_f(x) =  \log(x)$  that have singularities at $x=0$ .

Denote the capacity of link $l$ as $C_l$ and assume it is a fixed and positive constant.\footnote{As stated in \cite{LiuShroff15TON}, this is a suitable model for wireline networks and wireless networks with fixed transmission power and orthogonal channels.}  Define $\mu_{l}^{(f)} $ as the amount of  session $f$'s data routed at link $l$ that is to be determined by our algorithm. Note that in general, the network may be configured such that some session $f$ is forbidden to use link $l$.   For each link $l$, define $\mathcal{S}_{l}\subseteq \mathcal{F}$ as the set of sessions that are allowed to use link $l$.  The case of unrestricted routing 
is treated by defining $\mathcal{S}_l = \mathcal{F}$ for all links $l$. 

Note that if $l=(n,m)$ with $n,m\in \mathcal{N}$, then $\mu_{l}^{(f)}$ and $C_{l}$ can also be respectively written as $\mu_{(n,m)}^{(f)}$ and $C_{(n,m)}$.  For each node $n\in \mathcal{N}$, denote the sets of its incoming links  and outgoing links as $\mathcal{I}(n)$ and $\mathcal{O}(n)$, respectively. Note that $x_f,\forall f\in \mathcal{F}$ and $\mu_{l}^{(f)}, \forall l\in \mathcal{L}, \forall f\in \mathcal{F}$ are the decision variables of a joint rate control and routing algorithm.  If the global network topology information is available, the optimal joint rate control and routing can be formulated as the following multi-commodity network flow problem:
\begin{align}
\max_{x_f, \mu_{l}^{(f)}}  &\sum_{f\in \mathcal{F}} U_{f}(x_{f}) \label{eq:opt-obj}\\
\text{s.t.} \quad  &   x_{f} \mathbf{1}_{\{n=\text{Src}(f)\}} + \sum_{l\in \mathcal{I}(n)} \mu_{l}^{(f)} \leq   \sum_{l\in \mathcal{O}(n)} \mu_{l}^{(f)},\forall f\in \mathcal{F} , \forall n\in\mathcal{N}\setminus\{\text{Dst}(f)\}\label{eq:opt-flow-balance-cons}\\
			 &  \sum_{f\in \mathcal{F}} \mu_{l}^{(f)}  \leq C_l, \forall l\in\mathcal{L}, \label{eq:opt-link-capacity-cons}\\
			 &\mu_{l}^{(f)} \geq 0, \forall l\in \mathcal{L},\forall f\in  \mathcal{S}_{l},\label{eq:opt-allowed-link}\\
			 &\mu_{l}^{(f)} = 0, \forall l\in \mathcal{L},\forall f\in \mathcal{F}\setminus\mathcal{S}_{l} \label{eq:opt-forbiden-link},\\
			 &x_{f} \in \dom(U_f), \forall f\in \mathcal{F} \label{eq:opt-rate-nonnegative}
\end{align}
where $\mathbf{1}_{\{\cdot\}}$ is an indicator function; \eqref{eq:opt-flow-balance-cons} represents the node flow conservation constraints relaxed by replacing the equality with an inequality, meaning that the total rate of flow $f$ into node $n$ is less than or equal to the total rate of flow $f$ out of the node (since, in principle, we can always send fake data for departure links when the inequality is loose); and \eqref{eq:opt-link-capacity-cons} represents link capacity constraints. Note that for each flow $f$, there is no constraint \eqref{eq:opt-flow-balance-cons} at its destination node $\text{Dst}(f)$ since all incoming data are consumed by this node. 

The above formulation includes network utility maximization with fixed paths as special cases. In the case when each session only has one single given path, e.g., the network utility maximization problem considered in \cite{Low99TON}, we could modify the sets $\mathcal{S}_l$ used in  constraints \eqref{eq:opt-allowed-link} and \eqref{eq:opt-forbiden-link} to reflect this fact. For example, if link $l_{1}$ is only used for sessions $f_{1}$ and $f_{2}$, then $\mathcal{S}_{l_{1}} = \{f_{1},f_{2}\}$. Similarly, the case \cite{LinShroff06TAC}  where each flow is restricted to using links from a set of predefined paths can be treated by modifying the sets $\mathcal{S}_l$ accordingly. See Appendix \ref{sec:multipath-num} for more discussions.

The solution to problem \eqref{eq:opt-obj}-\eqref{eq:opt-rate-nonnegative}  
corresponds to the optimal joint rate control and routing.  However, to solve this 
convex program at a single computer, we need to know the global network topology and the solution is a centralized one, which is not practical for large data networks. As observed in \cite{Neely05JSAC,Eryilmaz06JSAC,LinShroff04CDC,LiuShroff15TON}, various versions of backpressure algorithms can be interpreted as distributed solutions to 
problem \eqref{eq:opt-obj}-\eqref{eq:opt-rate-nonnegative} from first order Lagrangian dual type methods.  

\begin{Assumption} \label{assumption:feasible} (Feasibility) Problem \eqref{eq:opt-obj}-\eqref{eq:opt-rate-nonnegative} 
has at least one optimal solution vector  $[x_f^\ast; \mu_l^{(f),\ast}]_{f \in \mathcal{F}, l \in \mathcal{L}}$.
\end{Assumption} 


\begin{Assumption} \label{ass:Lagrange} (Existence of Lagrange multipliers) Assume the convex 
program \eqref{eq:opt-obj}-\eqref{eq:opt-rate-nonnegative} has Lagrange multipliers attaining the strong duality.  Specifically, 
define convex set $\mathcal{C} = \{[x_f; \mu_l^{(f)}]_{f\in \mathcal{F}, l\in\mathcal{L}} :  \eqref{eq:opt-link-capacity-cons}\text{-}\eqref{eq:opt-rate-nonnegative}~\text{hold} \}$. Assume there exists a Lagrange multiplier vector $\boldsymbol{\lambda}^{\ast} = [\lambda_{n}^{(f),\ast}]_{f\in \mathcal{F}, n\in\mathcal{N}\setminus\{\text{Dst}(f)\}} \geq \mathbf{0}$ such that 
\begin{align*}
q(\boldsymbol{\lambda}^{\ast}) = \sup\{\eqref{eq:opt-obj}: \eqref{eq:opt-flow-balance-cons}\text{-}\eqref{eq:opt-rate-nonnegative}\}
\end{align*} 
where $q(\boldsymbol{\lambda}) = \sup_{[x_{f}; \mu_{l}^{(f)}]\in \mathcal{C}} \big\{\sum_{f\in \mathcal{F}} U_{f}(x_{f}) - \sum_{f\in\mathcal{F}} \sum_{n\in \mathcal{N}\setminus\{\text{Dst}(f)\}} \lambda_{n}^{(f)} \big[x_{f} \mathbf{1}_{\{n=\text{Src}(f)\}} + \sum_{l\in \mathcal{I}(n)} \mu_{l}^{(f)} - \sum_{l\in \mathcal{O}(n)} \mu_{l}^{(f)}\big] \big\}$ is the Lagrangian dual function of problem  \eqref{eq:opt-obj}-\eqref{eq:opt-rate-nonnegative} by treating \eqref{eq:opt-link-capacity-cons}-\eqref{eq:opt-rate-nonnegative} as a convex set constraint.
\end{Assumption} 

Assumptions \ref{assumption:feasible} and \ref{ass:Lagrange} hold in most cases of interest.  For example, Slater's condition guarantees Assumption \ref{ass:Lagrange}. Since the constraints \eqref{eq:opt-flow-balance-cons}-\eqref{eq:opt-rate-nonnegative} are linear, Proposition 6.4.2 
in  \cite{book_ConvexAnalysisOpt_Bertsekas} ensures that 
Lagrange multipliers exist whenever constraints  \eqref{eq:opt-flow-balance-cons}-\eqref{eq:opt-rate-nonnegative} are
feasible and when the utility functions $U_f$ are either defined over open sets (such as $U_f(x) = \log(x)$ with $\dom(U_f)=(0,\infty)$)
or can be \emph{concavely extended} to open sets, meaning that there is an $\epsilon>0$ and a concave  
function $\widetilde{U}_f:(-\epsilon, \infty)\rightarrow\mathbb{R}$ such that $\widetilde{U}_f(x)=U_f(x)$ whenever $x \geq 0$.\footnote{If $\dom(U_f)=[0,\infty)$, such concave 
extension is possible if the right-derivative 
of $U_f$ at $x=0$ is finite (such as for $U_f(x) = \log(1+x)$ or $U_f(x) = \min[x, 3]$).  Such an extension is impossible for the example $U_f(x)=\sqrt{x}$  because the slope is infinite at $x=0$.  Nevertheless, Lagrange multipliers often exist even for these utility functions, such as when Slater's condition holds \cite{book_ConvexAnalysisOpt_Bertsekas}.} 

\begin{Fact}\label{fact:optimal-solution-attain-equality} (Replacing inequality with equality) If Assumption \ref{assumption:feasible} holds, 
problem \eqref{eq:opt-obj}-\eqref{eq:opt-rate-nonnegative} has an optimal solution vector $[x_f^\ast; \mu_l^{(f),\ast}]_{f \in \mathcal{F}, l \in \mathcal{L}}$ such that all constraints \eqref{eq:opt-flow-balance-cons} take equalities.
\end{Fact}

\begin{IEEEproof}
Note that each $\mu_{l}^{(f)}$ can appear on the left side in at most one constraint \eqref{eq:opt-flow-balance-cons} and appear on the right side in at most one constraint \eqref{eq:opt-flow-balance-cons}. Let  $[x_f^\ast; \mu_l^{(f),\ast}]_{f \in \mathcal{F}, l \in \mathcal{L}}$ be an optimal solution vector such that at least one inequality constraint \eqref{eq:opt-flow-balance-cons} is loose. Note that we can reduce the value of $\mu_{l}^{(f),\ast}$ on the right side of a loose \eqref{eq:opt-flow-balance-cons} until either that constraint holds with equality, or until $\mu_{l}^{(f),\ast}$ reduces to $0$. The objective function value does not change, and no constraints are violated.  We can repeat the process until all inequality constraints \eqref{eq:opt-flow-balance-cons} are tight.
\end{IEEEproof}

\section{The New Backpresure Algorithm}

\subsection{Discussion of Various Queueing Models} \label{sec:queue-model}
At each node, an independent queue backlog is maintained for each session.  At each slot $t$, let $x_{f}[t]$ be the source session rates; and let $\mu_{l}^{(f)}[t]$ be the link session rates. Some prior work enforces the constraint \eqref{eq:opt-flow-balance-cons} via virtual queues $Y_{n}^{(f)}[t]$ of the following form: 
\begin{align}
Y_{n}^{(f)}[t+1] = &\max\Big\{ Y_{n}^{(f)}[t]  +x_{f}[t] \mathbf{1}_{\{n=\text{Src}(f)\}} + \sum_{l\in \mathcal{I}(n)} \mu_{l}^{(f)}[t]- \sum_{l\in \mathcal{O}(n)} \mu_{l}^{(f)}[t], 0\Big\}.    \label{eq:queue-update-model1}
\end{align}
While this virtual equation is a meaningful approximation, it differs from reality in that  new injected data are allowed to be transmitted immediately, or equivalently, a single packet is allowed to enter and leave many nodes within the same slot. Further, there is no clear connection between the virtual queues $Y_{n}^{(f)}[t]$ in \eqref{eq:queue-update-model1} and the actual queues in the network. Indeed, it is easy to construct
examples that show there can be an arbitrarily large difference between the $Y_{n}^{(f)}[t]$ value in \eqref{eq:queue-update-model1} and the physical queue size in actual networks (see Appendix \ref{sec:example-queue}).

An actual queueing network has queues $Z_{n}^{(f)}[t]$ with the following dynamics: 
\begin{align}
Z_{n}^{(f)}[t+1] \leq &\max\Big\{ Z_{n}^{(f)}[t]  - \sum_{l\in \mathcal{O}(n)} \mu_{l}^{(f)}[t], 0\Big\} +x_{f}[t] \mathbf{1}_{\{n=\text{Src}(f)\}} + \sum_{l\in \mathcal{I}(n)} \mu_{l}^{(f)}[t].    \label{eq:queue-update-model2}
\end{align}
This is faithful to actual queue dynamics and does not allow data to be retransmitted over multiple hops in one slot. Note that  \eqref{eq:queue-update-model2} is an inequality because the
new arrivals from other nodes may be strictly less than $\sum_{l\in \mathcal{I}(n)} \mu_{l}^{(f)}[t]$ because those other nodes may not have enough backlog to send. The model \eqref{eq:queue-update-model2} allows for any decisions to be made to fill the transmission values $\mu_l^{(f)}[t]$ in the case
that $Z_{n}^{(f)}[t] \leq \sum_{l\in \mathcal{O}(n)} \mu_{l}^{(f)}[t]$, provided that \eqref{eq:queue-update-model2} holds.

This paper develops an algorithm that converges to the optimal utility defined by problem \eqref{eq:opt-obj}-\eqref{eq:opt-rate-nonnegative}, and that produces worst-case bounded queues on the actual queueing network, that is, with actual queues that evolve as given in \eqref{eq:queue-update-model2}. To begin, it is convenient to introduce the following virtual queue equation
\begin{align}
Q_{n}^{(f)}[t+1] = &Q_{n}^{(f)}[t]   - \sum_{l\in \mathcal{O}(n)} \mu_{l}^{(f)}[t] +x_{f}[t] \mathbf{1}_{\{n=\text{Src}(f)\}} + \sum_{l\in \mathcal{I}(n)} \mu_{l}^{(f)}[t],  \label{eq:queue-update}
\end{align}
where $Q_{n}^{(f)}[t]$ represents a {\bf virtual queue} value associated with session $f$ at node
$n$. At first glance, this model \eqref{eq:queue-update} appears to be only an approximation, perhaps
even a  worse approximation than \eqref{eq:queue-update-model1}, because it allows the $Q_{n}^{(f)}[t]$ values to be negative. Indeed, we use $Q_{n}^{(f)}[t]$ only as virtual queues to inform the algorithm and do not treat them as actual queues. However, this paper shows that using
these virtual queues to choose the $\boldsymbol{\mu}[t]$ decisions ensures not only that the desired constraints \eqref{eq:opt-flow-balance-cons} are satisfied, but that the resulting $\boldsymbol{\mu}[t]$ decisions create bounded queues $Z_{n}^{(f)}[t]$ in the {\bf actual network}, where the actual queues evolve according to \eqref{eq:queue-update-model2}. In short, our algorithm can be faithfully implemented with respect to actual queueing networks, and converges to exact optimality on those networks.

The next lemma shows that if an algorithm can guarantee virtual queues $Q_{n}^{(f)}[t]$ defined in \eqref{eq:queue-update} are bounded, then actual physical queues satisfying \eqref{eq:queue-update-model2} are also bounded.

\begin{Lem}\label{lm:queue-model-relation}
Consider a network flow problem described by problem \eqref{eq:opt-obj}-\eqref{eq:opt-rate-nonnegative}.  For all $l \in \mathcal{L}$ and $f \in \mathcal{F}$, let 
$\mu_{l}^{(f)}[t], x_{f}[t]$ be decisions yielded by a dynamic algorithm.  Suppose $Y_{n}^{(f)}[t]$, $Z_{n}^{(f)}[t]$, $Q_n^{(f)}[t]$ 
evolve by \eqref{eq:queue-update-model1}-\eqref{eq:queue-update} with initial conditions 
$V_n^{(f)}[0]=Z_{n}^{(f)}[0]=Q_{n}^{(f)}[0] = 0$.  If there exists a constant $B >0$ such that $\vert Q_{n}^{(f)}[t]\vert \leq B, \forall t$, then
\begin{enumerate}
\item $Z_{n}^{(f)}[t] \leq 2B + \sum_{l\in \mathcal{O}(n)}C_{l}$ for all $t \in \{0, 1, 2, \ldots\}$.
\item $Y_{n}^{(f)}[t] \leq 2B + \sum_{l\in \mathcal{O}(n)} C_{l}$ for all $t \in \{0, 1, 2, \ldots\}$.
\end{enumerate} 
\end{Lem}
\begin{IEEEproof}
\begin{enumerate}
\item Fix $f\in \mathcal{F}, n\in \mathcal{N}\setminus\{\text{Dst}(f)\}$. Define an auxiliary virtual queue $\widehat{Q}_{n}^{(f)}[t]$ that is initialized by $\widehat{Q}_{n}^{(f)}[0] = B + \sum_{l\in \mathcal{O}(n)}C_{l}$ and evolves according to \eqref{eq:queue-update}. It follows that $\widehat{Q}_{n}^{(f)}[t] = Q_{n}^{(f)}[t] + B +  \sum_{l\in \mathcal{O}(n)}C_{l}, \forall t$. Since $Q_{n}^{(f)}[t] \geq -B, \forall t$ by assumption, we have $\widehat{Q}_{n}^{(f)}[t] \geq  \sum_{l\in \mathcal{O}(n)}C_{l} \geq \sum_{l\in \mathcal{O}(n)} \mu_{l}^{(f)}[t], \forall t$. This implies that  $\widehat{Q}_{n}^{(f)}[t]$ also satisfies:
\begin{align}
\widehat{Q}_{n}^{(f)}[t+1] = &\max\Big\{\widehat{Q}_{n}^{(f)}[t] - \sum_{l\in \mathcal{O}(n)} \mu_{l}^{(f)}[t], 0\Big\} +x_{f}[t] \mathbf{1}_{\{n=\text{Src}(f)\}} + \sum_{l\in \mathcal{I}(n)} \mu_{l}^{(f)}[t], \forall t \label{eq:pf-queue-model-relation-eq1}
\end{align} 
which is identical to \eqref{eq:queue-update-model2} except the inequality is replaced by an equality. Since $Z_{n}^{(f)}[0] =0 <\widehat{Q}_{n}^{(f)}[0]  $; and $\widehat{Q}_{n}^{(f)}[t]$ satisfies \eqref{eq:pf-queue-model-relation-eq1}, by inductions, $Z_{n}^{(f)}[t] \leq \widehat{Q}_{n}^{(f)}[t], \forall t$.

Since $\widehat{Q}_{n}^{(f)}[t] = Q_{n}^{(f)}[t] + B + \sum_{l\in \mathcal{O}(n)}C_{l}, \forall t$ and $Q_{n}^{(f)}[t]\leq B, \forall t$, we have $\widehat{Q}_{n}^{(f)}[t] \leq 2B + \sum_{l\in \mathcal{O}(n)}C_{l}, \forall t$. It follows that $Z_{n}^{(f)}[t] \leq 2B + \sum_{l\in \mathcal{O}(n)}C_{l}, \forall t$.
 
\item The proof of part (2) is similar and is in Appendix \ref{sec:proof-part2-lemma-ueue-model-relation}.

\end{enumerate}
\end{IEEEproof}

\subsection{The New Backpressure  Algorithm}

In this subsection, we propose a new backpressure algorithm that yields source session rates $x_f[t]$ and link session rates $\mu_l^{(f)}[t]$ at each slot such that the physical queues for each session at each node are bounded by a constant and the time average utility satisfies 
$$\frac{1}{t}\sum_{\tau=0}^{t-1} \sum_{f\in \mathcal{F}} U_{f}(x_{f}[t]) \geq \sum_{f\in \mathcal{F}} U_{f}(x_{f}^{\ast}) - O(1/t), \forall t$$ 
where $x_{f}^{\ast}$ are from the optimal solution to \eqref{eq:opt-obj}-\eqref{eq:opt-rate-nonnegative}. Note that Jensen's inequality further implies that 
$$\sum_{f\in \mathcal{F}} U_{f}\big(\frac{1}{t}\sum_{\tau=0}^{t-1}x_{f}[\tau]\big) \geq \sum_{f\in \mathcal{F}} U_{f}(x_{f}^{\ast}) - O(1/t), \forall t $$

The new backpressure algorithm is described in Algorithm \ref{alg:new-alg}. Similar to existing backpressure algorithms, the updates in Algorithm \ref{alg:new-alg} at each node $n$ are fully distributed and only depend on weights at itself and its neighbor nodes.  Unlike existing backpressure algorithms, the weights used to update decision variables 
$x_f[t]$ and $\mu_l^{(f)}[t]$ are not the virtual queues $Q_{n}^{(f)}[t]$ themselves,  rather, they 
are augmented values $W_n^{(f)}[t]$ equal to the sum of the virtual queues and the amount of net injected data in the previous slot $t-1$.  In addition,  the updates  involve an additional quadratic term, which is similar to a term used in proximal algorithms \cite{Parikh13ProximalAlgorithm}. 
\begin{algorithm}
\caption{The New Backpressure  Algorithm}
\label{alg:new-alg}
Let $\alpha_{n}>0, \forall n\in \mathcal{N}$ be constant parameters.  Initialize $x_{f}[-1] = 0$, $\mu_{l}^{(f)} [-1]= 0, \forall f\in \mathcal{F}, \forall l\in \mathcal{L}$ and $Q_{n}^{(f)}[0] = 0, \forall n\in \mathcal{N}, \forall f\in \mathcal{F}$. At each time $t\in\{0,1,2,\ldots\}$, each node $n$ does the following:
\begin{itemize}
\item For each $ f\in \mathcal{F}$, if node $n$ is not the destination node of session $f$, i.e., $n\neq \text{Dst}(f)$, then  define weight $W_n^{(f)}[t]$: 
\begin{align}			 
W_{n}^{(f)}[t] = &Q_{n}^{(f)}[t] +x_{f}[t-1] \mathbf{1}_{\{n=\text{Src}(f)\}} + \sum_{l\in \mathcal{I}(n)} \mu_{l}^{(f)}[t-1] - \sum_{l\in \mathcal{O}(n)} \mu_{l}^{(f)}[t-1], \label{eq:alg-weight-update} 
\end{align}
If node $n$ is the destination node, i.e., $n = \text{Dst}(f)$, then define $W_{n}^{(f)}[t] =0$.  Notify neighbor nodes (nodes $k$ that can send session $f$ to node 
$n$, i.e.,   
$\forall k$ such that $f \in \mathcal{S}_{(k,n)}$)  
about this new $W_n^{(f)}[t]$ value. 

\item For each $f\in \mathcal{F}$, if node $n$ is the source node of session $f$, i.e., $n=\text{Src}(f)$, choose $x_{f}[t]$ as the solution to 
\begin{align}
\max_{x_{f}} \quad &  U_{f}(x_{f}) - W_{n}^{(f)}[t] x_{f} - \alpha_{n}  \big(x_{f} -x_{f}[t-1]\big)^{2} \label{eq:source-opt-obj}\\
\text{s.t.} \quad  &  x_{f} \in \dom(U_f) \label{eq:souce-opt-cons}
\end{align}

\item For all $(n,m) \in \mathcal{O}(n)$, choose $\{\mu_{(n,m)}^{(f)}[t], \forall f\in \mathcal{F}\}$ as the solution to the following convex program:
\begin{align}
\max_{\mu_{(n,m)}^{(f)}} ~ &\mbox{$\displaystyle\sum_{f\in \mathcal{F}}  \big( W_{n}^{(f)}[t]  - W_{m}^{(f)}[t]\big) \mu_{(n,m)}^{(f)}$}\mbox{$\displaystyle-\big(\alpha_{n} + \alpha_{m}\big)\sum_{f\in\mathcal{F}} \big(\mu_{(n,m)}^{(f)} -\mu_{(n,m)}^{(f)}[t-1]\big)^{2}$} \label{eq:routing-opt-obj}\\
\text{s.t.}\quad &  \sum_{f\in \mathcal{F}} \mu_{(n,m)}^{(f)}  \leq C_{(n,m)}\\
			 &\mu_{(n,m)}^{(f)} \geq 0, \forall f\in \mathcal{S}_{(n,m)}  \\
			 &\mu_{(n,m)}^{(f)} =0, \forall f \not\in \mathcal{S}_{(n,m)} \label{eq:routing-opt-cons-zero}
\end{align}
\item For each $ f\in \mathcal{F}$, if node $n$ is not the destination of $f$, i.e., $n\neq \text{Dst}(f)$,  update virtual queue $Q_n^{(f)}[t+1]$ by \eqref{eq:queue-update}. 
\end{itemize} 
\end{algorithm}

\subsection{Almost Closed-Form Updates in Algorithm \ref{alg:new-alg}}

This subsection shows the decisions $x_{f}[t]$ and $\mu_{l}^{(f)}[t]$ in Algorithm \ref{alg:new-alg} have 
either closed-form solutions or ``almost'' closed-form solutions at each iteration $t$.

\begin{Lem} Let $\hat{x}_f \equiv x_f[t]$ denote the solution to \eqref{eq:source-opt-obj}-\eqref{eq:souce-opt-cons}. 
\begin{enumerate}
\item Suppose $\dom(U_f)=[0,\infty)$ and $U_{f}(x_{f})$ is differentiable. Let $h(x_{f}) = U_{f}^{\prime}(x_{f}) - 2\alpha_{n} x_{f} + 2\alpha_{n} x_{f}[t-1] - W_{n}^{(f)}[t]$. If $h(0)<0$, then $\hat{x}_f=0$; otherwise $\hat{x}_{f}$ is the root to the equation $h(x_{f})= 0$ and can be found by a bisection search. 
\item Suppose $\dom(U_f)=(0,\infty)$ and $U_f(x_f)=w_f\log(x_f)$ for some weight $w_f>0$. 
Then: 
\begin{align*}
\hat{x}_f &= \frac{2\alpha_{n} x_f[t-1]-W_n^{(f)}[t]}{4\alpha_n}+\frac{\sqrt{(W_n^{(f)}[t]-2\alpha_{n} x_f[t-1])^2 + 8\alpha_n w_f}}{4\alpha_{n}}
\end{align*} 
\end{enumerate} 
\end{Lem}
\begin{IEEEproof}
Omitted for brevity. 
\end{IEEEproof}

The problem \eqref{eq:routing-opt-obj}-\eqref{eq:routing-opt-cons-zero} can be represented as follows by eliminating $\mu_{(n,m)}^{(f)}, f\not\in \mathcal{S}_{(n,m)}$, completing the square and replacing maximization with minimization. (Note that $K = \vert \mathcal{S}_{(n,m)}\vert \leq \vert\mathcal{F}\vert$.)
\begin{align}
\min \quad &\frac{1}{2}\sum_{k=1}^{K}  (z_{k} - a_{k})^{2}\label{eq:simple-routing-opt-obj}\\
\text{s.t.} \quad  &  \sum_{k=1}^{K} z_{k}  \leq b \label{eq:simple-routing-opt-inequality}\\
			 & z_{k} \geq 0, \forall k\in\{1,2,\ldots,K\} \label{eq:simple-routing-opt-cons}
\end{align}
\begin{Lem} \label{lm:solution-routing-opt}
The solution to problem \eqref{eq:simple-routing-opt-obj}-\eqref{eq:simple-routing-opt-cons} is given by $z_{k}^{\ast} = \max\{0, a_{k}-\theta^{\ast}\}, \forall k\in\{1,2,\ldots,K\}$ where $\theta^{\ast}\geq 0$ can be found either by a bisection search (See Appendix \ref{sec:proof-solution-routing-opt}) or by Algorithm \ref{alg:routing-update} with complexity $O(K\log K)$.
\end{Lem}
\begin{IEEEproof}
A similar problem where \eqref{eq:simple-routing-opt-inequality} is replaced with an equality constraint in considered in \cite{Duchi08ICML}. The optimal solution to this quadratic program is characterized by its KKT condition and a corresponding algorithm can be developed to obtain its KKT point.  A complete proof is presented in Appendix \ref{sec:proof-solution-routing-opt}. 
\end{IEEEproof}

\begin{algorithm} 
\caption{Algorithm to solve problem \eqref{eq:simple-routing-opt-obj}-\eqref{eq:simple-routing-opt-cons}} 
\label{alg:routing-update}
\begin{enumerate}
\item Check if $\sum_{k=1}^{K} \max\{0,a_{k}\}\leq b$ holds. If yes, let $\theta^{\ast} = 0$ and $z_{k}^{\ast} =\max\{0,a_k\}, \forall k\in\{1,2,\ldots,K\}$ and terminate the algorithm; else, continue to the next step.
\item Sort all $a_k, \in\{1,2,\ldots,K\}$ in a decreasing order $\pi$ such that $a_{\pi(1)} \geq a_{\pi(2)} \geq \cdots \geq a_{\pi(K)}$. Define $S_{0}=0$.
\item For $k=1$ to $K$
\begin{itemize}
\item  Let $S_{k} = S_{k-1} + a_k$. Let $\theta^{\ast} = \frac{S_k -b }{k}$. 
\item  If $\theta^{\ast}\geq 0$, $a_{\pi(k)} - \theta^{\ast}>0$ and $a_{\pi(k+1)} - \theta^{\ast} \leq 0$, then terminate the loop; else, continue to the next iteration in the loop. 
\end{itemize}
\item Let $z_{k}^{\ast} =\max\{0,a_{k} - \theta^{\ast}\}, \forall k\in\{1,2,\ldots,K\}$ and terminate the algorithm.
\end{enumerate}
\end{algorithm}

Note that step (3) in Algorithm \ref{alg:routing-update} has complexity $O(K)$ and hence the overall complexity of Algorithm \ref{alg:routing-update} is dominated by the sorting step (2) with complexity $O(K\log(K))$.

\section{Performance Analysis of Algorithm \ref{alg:new-alg}}


\subsection{Basic Facts from Convex Analysis}

\begin{Def}[Lipschitz Continuity] \label{def:Lipschitz-continuous}
Let $\mathcal{Z} \subseteq \mathbb{R}^n$ be a convex set. Function $h: \mathcal{Z}\rightarrow \mathbb{R}^m$ is said to be Lipschitz continuous  on $\mathcal{Z}$ with modulus $\beta$ if there exists $\beta> 0$ such that $\Vert h(\mathbf{z}_{1}) - h(\mathbf{z}_{2}) \Vert \leq \beta \Vert\mathbf{z}_{1} - \mathbf{z}_{2}\Vert$  for all $ \mathbf{z}_{1}, \mathbf{z}_{2} \in \mathcal{Z}$. 
\end{Def}

\begin{Def}[Strongly Concave Functions]
 Let $\mathcal{Z} \subseteq \mathbb{R}^n$ be a convex set. Function $h$ is said to be strongly concave on $\mathcal{Z}$ with modulus $\alpha$ if there exists a constant $\alpha>0$ such that $h(\mathbf{z}) + \frac{1}{2} \alpha \Vert \mathbf{z} \Vert^2$ is concave on $\mathcal{Z}$.
\end{Def}

By the definition of strongly concave functions, it is easy to show that if $h(\mathbf{z})$ is concave and $\alpha>0$, then $h(\mathbf{z}) - \alpha \Vert \mathbf{z} - \mathbf{z}_0\Vert^2$ is strongly concave with modulus $2\alpha$ for any constant $\mathbf{z}_0$.

\begin{Lem} \label{lm:strong-convex-quadratic-optimality}
Let $\mathcal{Z} \subseteq \mathbb{R}^{n}$ be a convex set. Let function $h$ be strongly concave on $\mathcal{Z}$ with modulus $\alpha$ and $\mathbf{z}^{opt}$ be a global maximum of $h$ on $\mathcal{Z}$. Then, $h(\mathbf{z}^{opt}) \geq h(\mathbf{z}) + \frac{\alpha}{2} \Vert \mathbf{z}^{opt} - \mathbf{z}\Vert^{2}$ for all $\mathbf{z}\in \mathcal{Z}$.
\end{Lem}

\subsection{Preliminaries}

Define column vector $\mathbf{y}=[x_{f}; \mu_{l}^{(f)}]_{f\in \mathcal{F}, l\in \mathcal{L}}$. For each $f\in \mathcal{F}, n\in \mathcal{N}\setminus \{\text{Dst}(f)\}$, define column vector 
\begin{align}
\mathbf{y}_{n}^{(f)} = \left\{\begin{array}{ll} ~[x_{f}; \mu_{l}^{(f)}]_{l\in \mathcal{I}(n)\cup \mathcal{O}(n)} &\text{if}~ n=\text{Src}(f), \\\ [\mu_{l}^{(f)}]_{l\in \mathcal{I}(n)\cup \mathcal{O}(n)}&\text{else}, \end{array} \right. \label{eq:y-each-flow-balance-cons}
\end{align}
which is composed by the control actions appearing in each constraint \eqref{eq:opt-flow-balance-cons}; and introduce a function with respect to $\mathbf{y}_{n}^{(f)}$ as
\begin{align}
g_n^{(f)} (\mathbf{y}_{n}^{(f)} )= x_{f} \mathbf{1}_{\{n=\text{Src}(f)\}} + \sum_{l\in \mathcal{I}(n)} \mu_{l}^{(f)} -  \sum_{l\in \mathcal{O}(n)} \mu_{l}^{(f)} \label{eq:g-flow-balance-cons}
\end{align}
Thus, constraint \eqref{eq:opt-flow-balance-cons} can be rewritten as $$g_n^{(f)} (\mathbf{y}_{n}^{(f)} )\leq 0, \forall f\in \mathcal{F}, \forall n\in \mathcal{N}\setminus \{\text{Dst}(f)\}.$$
Note that each vector $\mathbf{y}_{n}^{(f)}$ is a subvector of $\mathbf{y}$ and has length $d_{n} +1$ where $d_{n}$ is the degree of node $n$ (the total number of outgoing links and incoming links) if node $n$ is the source of session $f$; and has length $d_{n}$ if node $n$ is not the source of session $f$. 
\begin{Fact}\label{fact:flow-cons-Lipschitz}
Each function $g_n^{(f)} (\cdot )$ defined in \eqref{eq:g-flow-balance-cons} is Lipschitz continuous with respect to vector $\mathbf{y}_{n}^{(f)}$ with modulus
\begin{align*}
\beta_{n} \leq  \sqrt{d_n+1}.
\end{align*}
where $d_n$ is the degree of node $n$.
\end{Fact}
\begin{IEEEproof}
This fact can be easily shown by noting that each $g_n^{(f)} (\mathbf{y}_{n}^{(f)} )$ is a linear function with respect to vector $\mathbf{y}_n^{(f)}$ and has at most $d_n+1$ non-zero coefficients that are equal to $\pm1$.
\end{IEEEproof}

Note that virtual queue update equation \eqref{eq:queue-update} can be rewritten as:
\begin{align} \label{eq:new-update} 
Q_{n}^{(f)}[t+1]=Q_{n}^{(f)}[t] + g_{n}^{(f)}(\mathbf{y}_{n}^{(f)}[t]), 
\end{align}
and  weight update equation \eqref{eq:alg-weight-update} can be rewritten as:
\begin{align}
W_{n}^{(f)}[t]=Q_{n}^{(f)}[t] + g_{n}^{(f)}(\mathbf{y}_{n}^{(f)}[t-1]) .
\end{align}
Define \begin{align}
L(t) = \frac{1}{2} \sum_{f\in \mathcal{F}}\sum_{n\in \mathcal{N}\setminus\text{Dst}(f)}\big(Q_{n}^{(f)}[t]\big)^{2} \label{eq:Lyapunov-fun}
\end{align}
and call it a {\it Lyapunov function}. In the remainder of this paper, double summations are often written compactly as  a single summation, e.g., 
\begin{align*}
\sum_{f\in \mathcal{F}}\sum_{n\in \mathcal{N}\setminus\text{Dst}(f)} \big(\cdot\big) \overset{\Delta}{=}\sum_{\begin{subarray}{c}f\in \mathcal{F},\\ n\in \mathcal{N}\setminus \text{Dst}(f)\end{subarray}} \big(\cdot\big).
\end{align*}
Define the Lyapunov drift as 
\begin{align*}
\Delta[t] = L(t+1) - L(t).
\end{align*}
The following lemma follows directly from equation \eqref{eq:new-update}.
\begin{Lem}
At each iteration $t\in \{0,1,\ldots\}$ in Algorithm \ref{alg:new-alg},  the Lyapunov drift is given by 
\begin{align}
\Delta[t] = \sum_{\begin{subarray}{c}f\in \mathcal{F},\\ n\in \mathcal{N}\setminus \text{Dst}(f)\end{subarray}}  \Big(Q_{n}^{(f)}[t] g_{n}^{(f)}(\mathbf{y}_{n}^{f}[t]) +\frac{1}{2}\big(g_{n}^{(f)}(\mathbf{y}_{n}^{f}[t])\big)^{2} \Big). \label{eq:drift-equality}
\end{align}
\end{Lem}
\begin{IEEEproof}
Fix $f\in \mathcal{F}$ and $n\in \mathcal{N}\setminus \text{Dst}(f)$, we have 
\begin{align}
&\frac{1}{2}\big(Q_{n}^{(f)}[t+1]\big)^{2} - \frac{1}{2}\big(Q_{n}^{(f)}[t]\big)^{2} \nonumber\\
\overset{(a)}{=}&  \frac{1}{2}\big(Q_{n}^{(f)}[t] + g_{n}^{(f)}(\mathbf{y}_{n}^{(f)}[t]) \big)^{2} - \frac{1}{2}\big(Q_{n}^{(f)}[t]\big)^{2}\nonumber \\
= & Q_{n}^{(f)}[t] g_{n}^{(f)}(\mathbf{y}_{n}^{f}[t]) +\frac{1}{2}\big(g_{n}^{(f)}(\mathbf{y}_{n}^{f}[t])\big)^{2}  \label{eq:pf-lm-drift-equality-eq1}
\end{align}
where (a) follows from \eqref{eq:new-update}.

By the definition of $\Delta[t]$, we have
\begin{align*}
\Delta[t] &=\frac{1}{2}\sum_{\begin{subarray}{c}f\in \mathcal{F},\\ n\in \mathcal{N}\setminus \text{Dst}(f)\end{subarray}}  \Big(\big(Q_{n}^{(f)}[t+1]\big)^{2} - \big(Q_{n}^{(f)}[t]\big)^{2}\Big)\\
&\overset{(a)}{=}\sum_{\begin{subarray}{c}f\in \mathcal{F},\\ n\in \mathcal{N}\setminus \text{Dst}(f)\end{subarray}}  \Big(Q_{n}^{(f)}[t] g_{n}^{(f)}(\mathbf{y}_{n}^{f}[t]) +\frac{1}{2}\big(g_{n}^{(f)}(\mathbf{y}_{n}^{f}[t])\big)^{2} \Big)
\end{align*}
where (a) follows from \eqref{eq:pf-lm-drift-equality-eq1}.
\end{IEEEproof}

Define $f(\mathbf{y}) = \sum_{f\in \mathcal{F}} U_{f}(x_{f})$.  At each time  $t$, consider choosing a decision vector 
$\mathbf{y}[t]$ that includes elements in each subvector $\mathbf{y}_{n}^{(f)}[t]$ to solve the following problem: 
\begin{align}
\max_{\mathbf{y}}~&f(\mathbf{y}) -\sum_{\begin{subarray}{c}f\in \mathcal{F},\\ n\in \mathcal{N}\setminus \text{Dst}(f)\end{subarray}} \big(W_{n}^{(f)}[t] g_{n}^{(f)}(\mathbf{y}_{n}^{(f)}) + \alpha_{n} \Vert \mathbf{y}_{n}^{(f)} - \mathbf{y}_{n}^{(f)}[t-1]\Vert^{2} \big)-\sum_{\begin{subarray}{c}f\in \mathcal{F},\\ n=\text{Dst}(f)\end{subarray}}\alpha_{n} \sum_{l\in \mathcal{I}(n)} ( \mu_l^{(f)} -\mu_{l}^{(f)}[t-1])^{2} \label{eq:alg-opt-obj}\\
\text{s.t.} ~~& \eqref{eq:opt-link-capacity-cons}\text{-}\eqref{eq:opt-rate-nonnegative} \label{eq:alg-opt-cons}
\end{align}  
The expression \eqref{eq:alg-opt-obj} is a \emph{modified drift-plus-penalty expression}. Unlike the standard drift-plus-penalty expressions from \cite{book_Neely10}, the above expression uses weights $W_{n}^{(f)}[t]$, which arguments each $Q_{n}^{(f)}[t]$ by $g_{n}^{(f)}(\mathbf{y}_{n}^{(f)}[t-1])$, rather than virtual queues $Q_{n}^{(f)}[t]$.  It also includes a ``prox''-like term that penalizes deviation from the previous $\mathbf{y}[t-1]$ vector. 
This results in the novel backpressure-type  algorithm of Algorithm \ref{alg:new-alg}.   Indeed, the decisions in 
Algorithm \ref{alg:new-alg} were \emph{derived} as the solution to the above problem \eqref{eq:alg-opt-obj}-\eqref{eq:alg-opt-cons}. This is formalized in the next lemma. 

\begin{Lem}\label{lm:alg-primal-opt}
At each iteration $t\in\{0,1,\ldots\}$, the action $\mathbf{y}[t]$ 
jointly chosen in Algorithm \ref{alg:new-alg} is the solution to problem \eqref{eq:alg-opt-obj}-\eqref{eq:alg-opt-cons}. 
\end{Lem}
\begin{IEEEproof}
The proof involves collecting terms associated with the $x_f[t]$ and $\mu_l^{(f)}[t]$ decisions. See Appendix \ref{sec:proof-lemma-alg-primal-opt} for details. 
\end{IEEEproof}

Furthermore, the next lemma summarizes that the action $\mathbf{y}[t]$ jointly chosen in Algorithm \ref{alg:new-alg}  provides a lower bound for the drift-plus-penalty expression at each iteration $t\in\{0,1,\ldots\}$.
\begin{Lem}\label{lm:dpp-bound}
Let $\mathbf{y}^{\ast} =[x_{f}^{\ast}; \mu_{l}^{(f),\ast}]_{f\in \mathcal{F}, l\in \mathcal{L}}$ be an optimal solution to problem \eqref{eq:opt-obj}-\eqref{eq:opt-rate-nonnegative} given in Fact \ref{fact:optimal-solution-attain-equality}, i.e., $g_{n}^{(f)}(\mathbf{y}_{n}^{(f),\ast}) = 0, \forall f\in \mathcal{F}, \forall n\in \mathcal{N}\setminus\text{Dst}(f)$. If $\alpha_{n} \geq  \frac{1}{2}(d_{n}+1), \forall n\in \mathcal{N}$, where $d_{n}$ is the degree of node $n$, then the action $\mathbf{y}[t] =[x_{f}[t]; \mu_{l}^{(f)}[t]]_{f\in \mathcal{F}, l\in \mathcal{L}}$ jointly chosen in Algorithm \ref{alg:new-alg} at each iteration $t\in\{0,1,\ldots\}$ satisfies
\begin{align*}
-\Delta[t] + f(\mathbf{y}[t]) \geq f(\mathbf{y}^{\ast}) + \Phi[t] - \Phi[t-1]
\end{align*} 
where $\Phi[t] =\sum_{f\in \mathcal{F}, n\in \mathcal{N}} \big(\alpha_{n}\mathbf{1}_{\{n\neq \text{Dst}(f)\}} \Vert \mathbf{y}_{n}^{(f),\ast} - \mathbf{y}_{n}^{(f)}[t]\Vert^{2}  + \alpha_{n} \mathbf{1}_{\{n=\text{Dst}(f)\}}\sum_{l\in \mathcal{I}(n)}( \mu_l^{(f),\ast} -\mu_{l}^{(f)}[t])^{2} \big)$.
\end{Lem}
\begin{IEEEproof}
See Appendix \ref{sec:proof-lemma-dpp-bond}.
\end{IEEEproof}

It remains to show that this modified backpressure algorithm leads to fundamentally improved performance. 

\subsection{Utility Optimality Gap Analysis}
Define column vector $\mathbf{Q}[t] = \big[Q_{n}^{(f)}[t]\big]_{f\in \mathcal{F}, n\in \mathcal{N}\setminus\{\text{Dst}(f)\}}$ as the stacked vector of all virtual queues $Q_{n}^{(f)}[t]$ defined in \eqref{eq:queue-update}. Note that  \eqref{eq:Lyapunov-fun} can be rewritten as $L(t) = \frac{1}{2} \Vert \mathbf{Q}[t]\Vert^2$. Define vectorized constraints \eqref{eq:opt-flow-balance-cons} as  $\mathbf{g}(\mathbf{y}) = [g_n^{(f)}(\mathbf{y}_{n}^{(f)})]_{f\in \mathcal{F}, n\in \mathcal{N}\setminus \text{Dst}(f)}$.

\begin{Lem} \label{lm:utility-inequality}
Let $\mathbf{y}^{\ast} =[x_{f}^{\ast}; \mu_{l}^{(f),\ast}]_{f\in \mathcal{F}, l\in \mathcal{L}}$ be an optimal solution to problem \eqref{eq:opt-obj}-\eqref{eq:opt-rate-nonnegative} given in Fact \ref{fact:optimal-solution-attain-equality}, i.e., $g_{n}^{(f)}(\mathbf{y}_{n}^{(f),\ast}) = 0, \forall f\in \mathcal{F}, \forall n\in \mathcal{N}\setminus\text{Dst}(f)$. If $\alpha_{n} \geq  \frac{1}{2}(d_{n}+1), \forall n\in \mathcal{N}$ in Algorithm \ref{alg:new-alg}, where $d_{n}$ is the degree of node $n$, then for all $t\geq 1$, 
\begin{align*}
\sum_{\tau=0}^{t-1} f(\mathbf{y}[\tau]) \geq t f(\mathbf{y}^{\ast}) - \zeta+ \frac{1}{2} \Vert \mathbf{Q}[t]\Vert^{2}.
\end{align*}
where $\zeta = \Phi[-1] = \sum_{f\in \mathcal{F}, n\in \mathcal{N}} \big(\alpha_{n}\mathbf{1}_{\{n\neq \text{Dst}(f)\}} \Vert \mathbf{y}_{n}^{(f),\ast}\Vert^{2}  + \alpha_{n} \mathbf{1}_{\{n=\text{Dst}(f)\}}\sum_{l\in \mathcal{I}(n)}( \mu_l^{(f),\ast})^{2} \big)$ is a constant.
\end{Lem}
\begin{IEEEproof}
By Lemma \ref{lm:dpp-bound}, we have $-\Delta[\tau] + f(\mathbf{y}[\tau]) \geq f(\mathbf{y}^{\ast})+ \Phi[t] - \Phi[t-1], \forall \tau\in\{0,1,\ldots,t-1\}$.  Summing over $\tau\in\{0,1,\ldots,t-1\}$ yields
\begin{align*}
&\sum_{\tau=0}^{t-1} f(\mathbf{y}[\tau]) -\sum_{\tau=0}^{t-1} \Delta[\tau] \\
\geq& t f(\mathbf{y}^{\ast}) +  \sum_{\tau=0}^{t-1} \big( \Phi[\tau] - \Phi[\tau-1]\big) \\
=& t f(\mathbf{y}^{\ast}) + \big(\Phi[t] - \Phi[-1]\big)\\
\overset{(a)}{\geq}&  t f(\mathbf{y}^{\ast}) -\Phi[-1]
\end{align*}
where (a) follows from the fact that $\Phi[t] \geq 0, \forall t$.

Recall  $\Delta[\tau] = L[\tau+1] - L[\tau] $, simplifying summations and rearranging terms yields
\begin{align*}
\sum_{\tau=0}^{t-1} f(\mathbf{y}[\tau]) \geq & t f(\mathbf{y}^{\ast}) - \Phi[-1] + L[t] - L[0]\\
\overset{(a)}{=}&t f(\mathbf{y}^{\ast}) - \Phi[-1] + \frac{1}{2} \Vert \mathbf{Q}[t]\Vert^{2} 
\end{align*}
where (a) follows from the fact that $L[0] = \mathbf{0}$ and $L[t] = \frac{1}{2} \Vert \mathbf{Q}[t]\Vert^2$.
\end{IEEEproof}

The next theorem summarizes that Algorithm \ref{alg:new-alg} yields a vanishing utility optimality gap that approaches  zero like $O(1/t)$.
\begin{Thm} \label{thm:utlity}
Let $\mathbf{y}^{\ast} =[x_{f}^{\ast}; \mu_{l}^{(f),\ast}]_{f\in \mathcal{F}, l\in \mathcal{L}}$ be an optimal solution to problem \eqref{eq:opt-obj}-\eqref{eq:opt-rate-nonnegative} given in Fact \ref{fact:optimal-solution-attain-equality}, i.e., $g_{n}^{(f)}(\mathbf{y}_{n}^{(f),\ast}) = 0, \forall f\in \mathcal{F}, \forall n\in \mathcal{N}\setminus\text{Dst}(f)$. If $\alpha_{n} \geq  \frac{1}{2}(d_{n}+1), \forall n\in \mathcal{N}$ in Algorithm \ref{alg:new-alg}, where $d_{n}$ is the degree of node $n$, then for all $t\geq 1$, we have
\begin{align*}
\frac{1}{t}\sum_{\tau=0}^{t-1} \sum_{f\in \mathcal{F}} U_{f}(x_{f}[\tau]) \geq \sum_{f\in \mathcal{F}} U_{f}(x_{f}^{\ast}) - \frac{1}{t} \zeta,
\end{align*}
where $\zeta$ is a constant defined in Lemma \ref{lm:utility-inequality}. Moreover, if we define $\overline{x}_{f}[t] = \frac{1}{t} \sum_{\tau=0}^{t-1} x_{f}[\tau], \forall f\in \mathcal{F}$, then 
\begin{align*}
\sum_{f\in \mathcal{F}} U_{f}(\overline{x}_{f}[t]) \geq \sum_{f\in \mathcal{F}} U_{f}(x_{f}^{\ast}) - \frac{1}{t} \zeta.
\end{align*}
\end{Thm}
\begin{IEEEproof}
Recall that $f(\mathbf{y}) = \sum_{f\in \mathcal{F}} U_{f}(x_{f})$. By Lemma \ref{lm:utility-inequality}, we have 
\begin{align*}
\sum_{\tau=0}^{t-1}\sum_{f\in \mathcal{F}} U_{f}(x_{f}[\tau]) \geq &t\sum_{f\in \mathcal{F}} U_{f}(x_{f}^{\ast}) -  \zeta + \frac{1}{2} \Vert \mathbf{Q}[t]\Vert^{2}\\
\overset{(a)}{\geq} &t\sum_{f\in \mathcal{F}} U_{f}(x_{f}^{\ast}) - \zeta .
\end{align*}
where (a) follows from the trivial fact that $\Vert \mathbf{Q}[t]\Vert^{2} \geq 0$.

Dividing both sides by a factor $t$ yields the first inequality in this theorem. The second inequality follows from the concavity of $U_{f}(\cdot)$ and Jensen's inequality.
\end{IEEEproof}

\subsection{Queue Stability Analysis} \label{sec:queue-length}
\begin{Lem}\label{lm:queue-g-equality}
Let $\mathbf{Q}[t], t\in\{0,1,\ldots\}$ be the virtual queues in Algorithm \ref{alg:new-alg}. For any $t\geq1$, 
\begin{align*}
\mathbf{Q}[t] = \sum_{\tau=0}^{t-1} \mathbf{g}(\mathbf{y}[\tau])
\end{align*} 
\end{Lem}
\begin{IEEEproof}
This lemma follows directly from the fact that $\mathbf{Q}[0] = \mathbf{0}$ and queue update equation \eqref{eq:queue-update} can be written as $\mathbf{Q}[t+1] = \mathbf{Q}[t] + \mathbf{g}(\mathbf{y}[t-1])$.
\end{IEEEproof}

The next theorem shows the boundedness of all virtual queues $Q_{n}^{(f)}[t]$ in Algorithm \ref{alg:new-alg}.
\begin{Thm}\label{thm:queue-bound}
Let $\mathbf{y}^{\ast} =[x_{f}^{\ast}; \mu_{l}^{(f),\ast}]_{f\in \mathcal{F}, l\in \mathcal{L}}$ be an optimal solution to problem \eqref{eq:opt-obj}-\eqref{eq:opt-rate-nonnegative} given in Fact \ref{fact:optimal-solution-attain-equality}, i.e., $g_{n}^{(f)}(\mathbf{y}_{n}^{(f),\ast}) = 0, \forall f\in \mathcal{F}, \forall n\in \mathcal{N}\setminus\text{Dst}(f)$, and $\boldsymbol{\lambda}^{\ast}$ be a Lagrange multiplier vector given in Assumption \ref{ass:Lagrange}. If $\alpha_{n} \geq  \frac{1}{2}(d_{n}+1)^2, \forall n\in \mathcal{N}$ in Algorithm \ref{alg:new-alg}, where $d_{n}$ is the degree of node $n$, then for all $t\geq 1$,  
\begin{align*}
\vert Q_{n}^{(f)}[t] \vert \leq  2\Vert \boldsymbol{\lambda}^{\ast}\Vert + \sqrt{2\zeta}, \forall f\in \mathcal{F}, \forall n\in \mathcal{N}\setminus \{\text{Dst}(f)\}.
\end{align*}
where $\zeta$ is a constant defined in Lemma \ref{lm:utility-inequality}.
\end{Thm}
\begin{IEEEproof}
Let $q(\boldsymbol{\lambda}) = \sup_{\mathbf{y}\in \mathcal{C}} \big\{f(\mathbf{y}) - \boldsymbol{\lambda}\tran \mathbf{g}(\mathbf{y})\big\}$ be the Lagrangian dual function defined in Assumption \ref{ass:Lagrange}. For all $\tau\in\{0,1,\ldots, \}$, by Assumption \ref{ass:Lagrange}, we have
\begin{align*}
f(\mathbf{y}^{\ast}) = q(\boldsymbol{\lambda}^{\ast}) \overset{(a)}{\geq} f(\mathbf{y}[\tau]) - \boldsymbol{\lambda}^{\ast,\tranT}\mathbf{g}(\mathbf{y}[\tau])
\end{align*}
where $(a)$ follows from the definition of $q(\boldsymbol{\lambda}^{\ast})$. Rearranging terms yields
\begin{align*}
 f(\mathbf{y}[\tau]) \leq f(\mathbf{y}^{\ast}) + \boldsymbol{\lambda}^{\ast,\tranT}\mathbf{g}(\mathbf{y}[\tau]), \forall \tau\in\{0,1,\ldots\}.
\end{align*}
Fix $t>0$. Summing over $\tau\in\{0,1,\ldots, t-1\}$ yields
\begin{align*}
\sum_{\tau=0}^{t-1}  f(\mathbf{y}[\tau]) \leq &t f(\mathbf{y}^{\ast}) + \sum_{\tau=0}^{t-1}\boldsymbol{\lambda}^{\ast,\tranT}\mathbf{g}(\mathbf{y}[\tau])\\
=&t f(\mathbf{y}^{\ast}) + \boldsymbol{\lambda}^{\ast,\tranT}\sum_{\tau=0}^{t-1}\mathbf{g}(\mathbf{y}[\tau])\\
\overset{(a)}{=}&t f(\mathbf{y}^{\ast}) + \boldsymbol{\lambda}^{\ast,\tranT}\mathbf{Q}[t]\\
\overset{(b)}{\leq}& tf(\mathbf{y}^{\ast}) + \Vert \boldsymbol{\lambda}^{\ast}\Vert \Vert\mathbf{Q}[t]\Vert
\end{align*}
where (a) follows form Lemma \ref{lm:queue-g-equality} and (b) follows from Cauchy-Schwarz inequality.

On the other hand, by Lemma \ref{lm:utility-inequality}, we have
\begin{align*}
\sum_{\tau=0}^{t-1} f(\mathbf{y}[\tau]) \geq tf(\mathbf{y}^{\ast}) - \zeta + \frac{1}{2} \Vert \mathbf{Q}[t]\Vert^{2}.
\end{align*}
Combining the last two inequalities and cancelling the common terms yields
\begin{align*}
&\frac{1}{2} \Vert \mathbf{Q}[t]\Vert^{2} - \zeta \leq \Vert \boldsymbol{\lambda}^{\ast}\Vert \Vert\mathbf{Q}[t]\Vert\\
\Rightarrow & \big( \Vert \mathbf{Q}[t]\Vert - \Vert \boldsymbol{\lambda}^{\ast}\Vert\big)^{2} \leq \Vert\boldsymbol{\lambda}^{\ast}\Vert^{2} + 2\zeta \\
\Rightarrow & \Vert \mathbf{Q}[t]\Vert \leq \Vert \boldsymbol{\lambda}^{\ast}\Vert + \sqrt{\Vert\boldsymbol{\lambda}^{\ast}\Vert^{2} + 2\zeta}\\
\overset{(a)}{\Rightarrow} & \Vert \mathbf{Q}[t]\Vert \leq 2\Vert \boldsymbol{\lambda}^{\ast}\Vert + \sqrt{ 2\zeta}
\end{align*}
where (a) follows from the basic inequality $\sqrt{a+b}\leq \sqrt{a} + \sqrt{b}$ for any $a,b\geq 0$.

Thus, for any $f\in \mathcal{F}$ and $n\in \mathcal{N}\setminus \{\text{Dst}(f)\}$, we have
\begin{align*}
\vert Q_{n}^{(f)}[t] \vert \leq  \Vert \mathbf{Q}[t]\Vert \leq 2\Vert \boldsymbol{\lambda}^{\ast}\Vert + \sqrt{ 2\zeta}.
\end{align*}
\end{IEEEproof}

This theorem shows that the absolute values of all virtual queues $Q_{n}^{(f)}[t]$ are bounded by a constant $B =  2\Vert \boldsymbol{\lambda}^{\ast}\Vert + \sqrt{ 2\zeta}$ from above. By Lemma \ref{lm:queue-model-relation} and discussions in Section \ref{sec:queue-model}, the actual physical queues $Z_{n}^{(f)}[t]$ evolving via \eqref{eq:queue-update-model2} satisfy $Z_{n}^{(f)}[t] \leq 2B+ \sum_{l\in \mathcal{O}(n)}C_{l}, \forall t$. This is summarized in the next corollary.

\begin{Cor}\label{cor:physical-queue-bound}
Let $\mathbf{y}^{\ast} =[x_{f}^{\ast}; \mu_{l}^{(f),\ast}]_{f\in \mathcal{F}, l\in \mathcal{L}}$ be an optimal solution to problem \eqref{eq:opt-obj}-\eqref{eq:opt-rate-nonnegative} given in Fact \ref{fact:optimal-solution-attain-equality}, i.e., $g_{n}^{(f)}(\mathbf{y}_{n}^{(f),\ast}) = 0, \forall f\in \mathcal{F}, \forall n\in \mathcal{N}\setminus\text{Dst}(f)$, and $\boldsymbol{\lambda}^{\ast}$ be a Lagrange multiplier vector given in Assumption \ref{ass:Lagrange}. If $\alpha_{n} \geq  \frac{1}{2}(d_{n}+1)^2, \forall n\in \mathcal{N}$ in Algorithm \ref{alg:new-alg}, where $d_{n}$ is the degree of node $n$, then all actual physical queues $Z_{n}^{(f)}[t], \forall f\in\mathcal{F}, \forall n\in \mathcal{N}\setminus \{\text{Dst}(f)\}$ in the network evolving via \eqref{eq:queue-update-model2} satisfy
\begin{align*}
Z_{n}^{(f)}[t] \leq &4\Vert \boldsymbol{\lambda}^{\ast}\Vert + 2\sqrt{ 2\zeta}+ \sum_{l\in \mathcal{O}(n)}C_{l}, \quad \forall t.
\end{align*}
where $\zeta$ is a constant defined in Lemma \ref{lm:utility-inequality}.
\end{Cor}

\subsection{Performance of Algorithm \ref{alg:new-alg}}
Theorems \ref{thm:utlity} and \ref{thm:queue-bound} together imply that Algorithm \ref{alg:new-alg} with $\alpha_{n} \geq  \frac{1}{2}(d_{n}+1), \forall n\in \mathcal{N}$ can achieve a vanishing utility optimality gap that decays like $O(1/t)$, where $t$ is number of iterations, and guarantees the physical queues at each node for each session are always bounded by a constant that is independent of the utility optimality gap. 

This is superior to existing backpressure algorithms from \cite{Eryilmaz06JSAC,book_Neely10,LiuShroff15TON} that can achieve an $O(1/V)$ utility gap only at the cost of an $O(V^2)$ or $O(V)$ queue length, where $V$ is an algorithm parameter. To obtain a vanishing utility gap, existing backpressure algorithms in \cite{Eryilmaz06JSAC, book_Neely10,LiuShroff15TON} necessarily yield unbounded queues. To obtain a vanishing utility gap, existing backpressure algorithms in \cite{Eryilmaz06JSAC, book_Neely10} yield unbounded queues. We also comment that $O(V^2)$ queue bound in the primal-dual type backpressure algorithm  \cite{Eryilmaz06JSAC} is actually of the order $V^2\Vert \boldsymbol{\lambda}^{\ast}\Vert +B_{1}$ where $\boldsymbol{\lambda}^{\ast}$ is the Lagrangian multiplier vector attaining strong duality and $B_{1}$ is a constant determined by the problem parameters.  A recent work \cite{Neely14Arxiv_ConvergenceTime} also shows that the $O(V)$ queue bound in the backpressure algorithm from drift-plus-penalty is of the order $V\Vert \boldsymbol{\lambda}^{\ast}\Vert +B_{2}$ where $B_{2}$ is also a constant determined by the problem parameters.  Since $\boldsymbol{\lambda}^{\ast}$ is a constant vector independent of $V$, both algorithms are claimed to have $O(V^2)$ or $O(V)$ queue bounds.  By Corollary \ref{cor:physical-queue-bound}, Algorithm \ref{alg:new-alg} guarantees physical queues at each node are bounded by  $4\Vert \boldsymbol{\lambda}^{\ast}\Vert + B_3$, where $B_3$ is constant given a problem.  Thus,  the constant queue bound guaranteed by Algorithm \ref{alg:new-alg} is typically smaller than the $O(V^2)$ or $O(V)$ queue bounds from  \cite{Eryilmaz06JSAC} and \cite{Neely14Arxiv_ConvergenceTime} even for a small $V$. (A small $V$ can yield a poor utility performance in the backpressure algorithms in \cite{Eryilmaz06JSAC, book_Neely10}.)

\section{Numerical Experiment}
\begin{figure}[htbp]
\centering
   \includegraphics[width=0.7\textwidth,height=0.7\textheight,keepaspectratio=true]{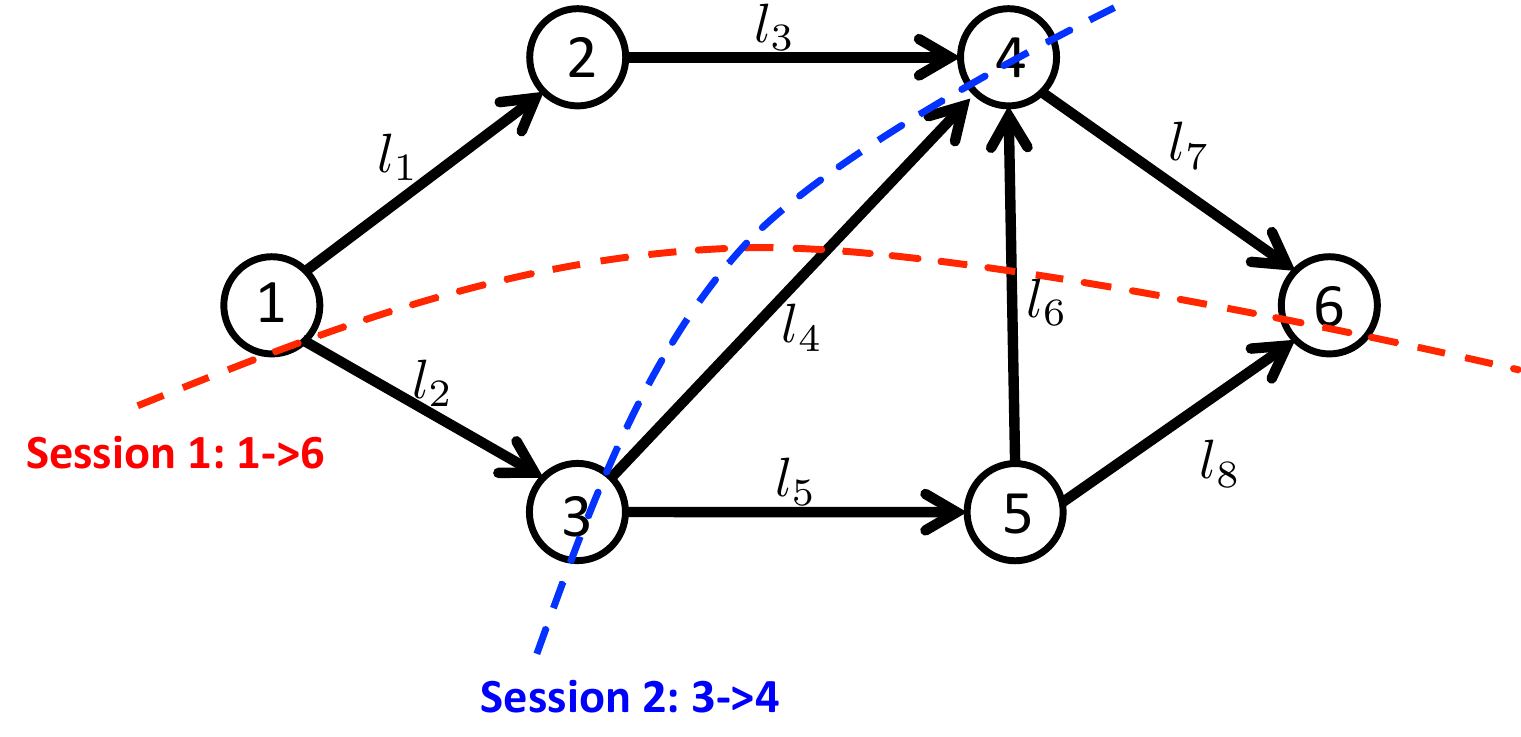} 
   \caption{A simple network with $6$ nodes, $8$ links and $2$ sessions.}
   \label{fig:network}
\end{figure}

\begin{figure}[htbp]
\centering
   \includegraphics[width=0.7\textwidth,height=0.7\textheight,keepaspectratio=true]{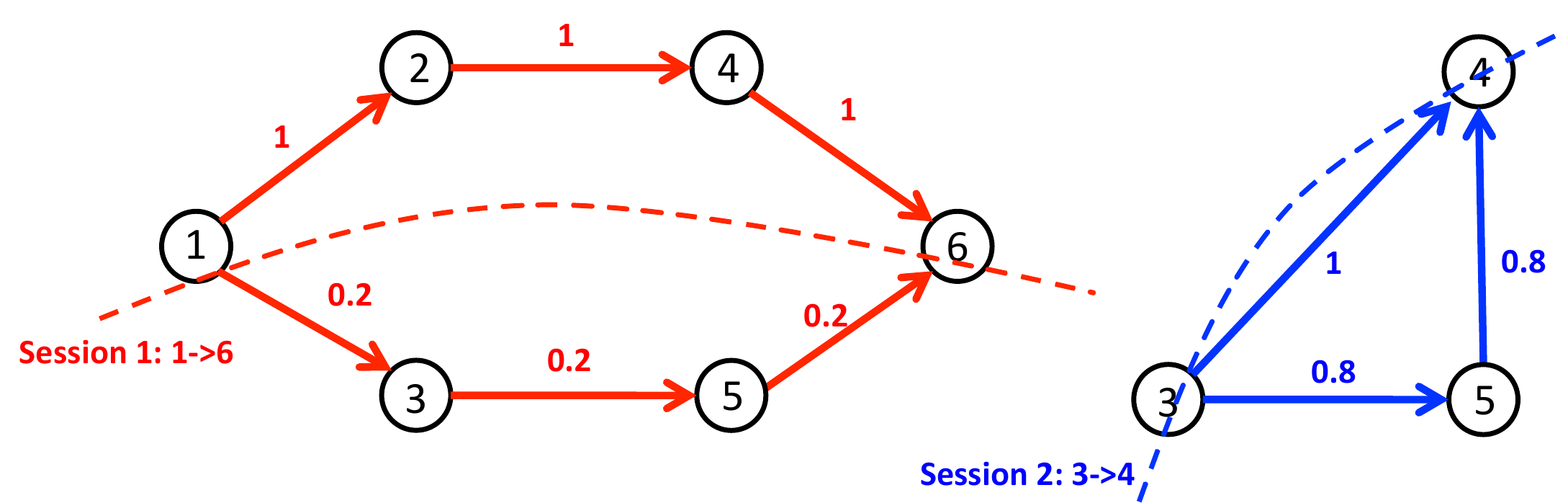} 
   \caption{The optimal routing for the network in Figure \ref{fig:network}.}
   \label{fig:routing-sol}
\end{figure}

In this section, we consider a simple network with $6$ nodes and $8$ links and $2$ sessions as described in Figure \ref{fig:network}.  This network has two sessions:  session $1$ from node $1$ to node $6$ has utility function $\log(x_{1})$ and session $2$ from node $3$ to node $4$ has utility function $1.5\log(x_{2})$.  (The log utilities are widely used as metrics of proportional fairness in the network \cite{Kelly98JORS}.) The routing path of each session is arbitrary as long as data can be delivered from the source node to the destination node.  For simplicity, assume that each link has capacity $1$. The optimal  source session rate to problem \eqref{eq:opt-obj}-\eqref{eq:opt-rate-nonnegative} is $x^{\ast}_{1} = 1.2$ and $x^{\ast}_{2} = 1.8$ and link session rates, i.e., static routing for each session, is drawn in Figure \ref{fig:routing-sol}. 

\begin{figure}[htbp]
\centering
   \includegraphics[width=0.7\textwidth,height=0.7\textheight,keepaspectratio=true]{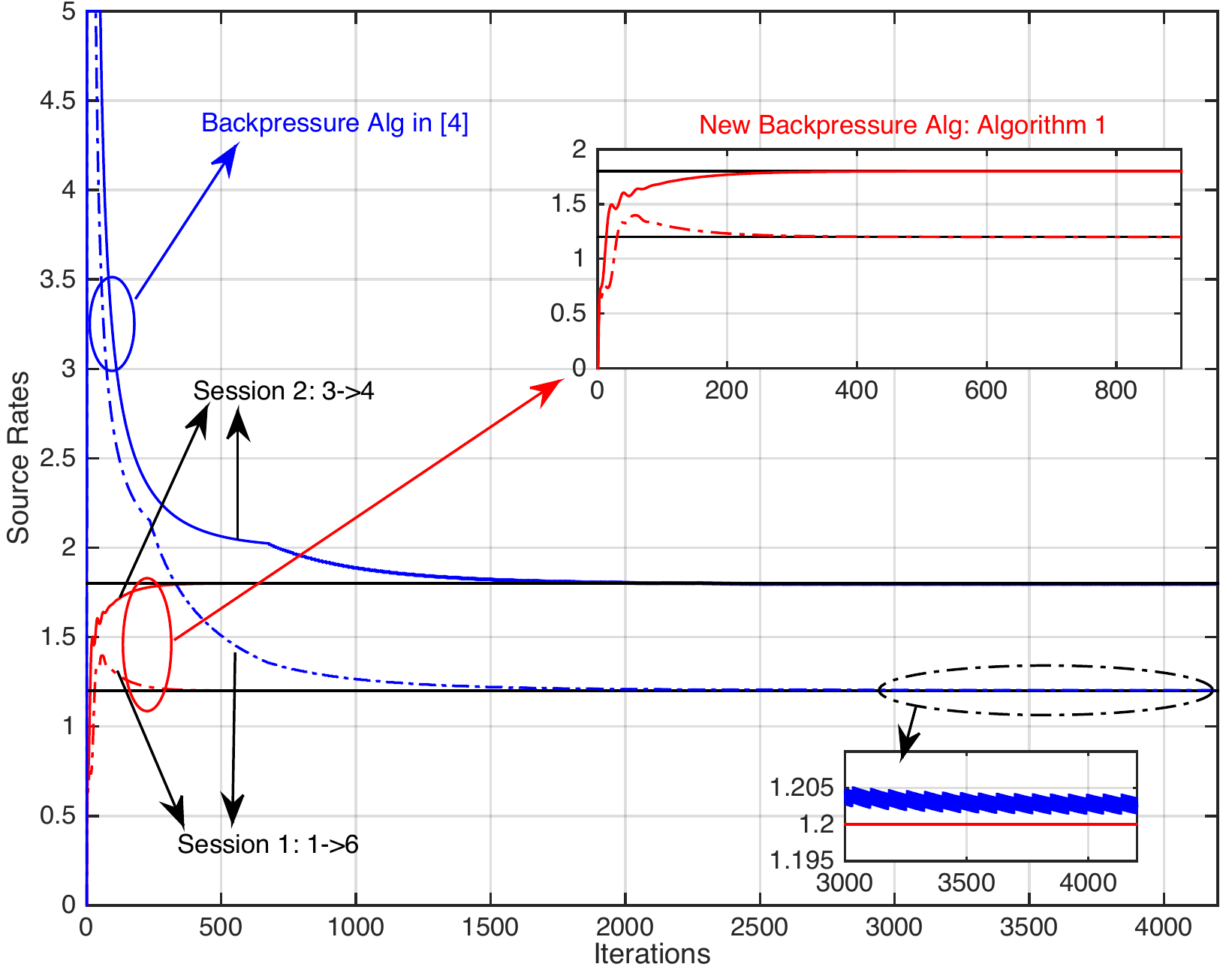} 
   \caption{Convergence performance comparison between Algorithm \ref{alg:new-alg} and the backpressure algorithm in \cite{book_Neely10}.}
   \label{fig:source-rate-convergence}
\end{figure}

To compare the convergence  performance of Algorithm \ref{alg:new-alg} and the backpressure algorithm in \cite{book_Neely10} (with the best utility-delay tradeoff among all existing backpressure algorithms), we run both Algorithm \ref{alg:new-alg} with $\alpha_{n} = \frac{1}{2}\big(d_{n}+1), \forall n\in \mathcal{N}$ and the backpressure algorithm in  \cite{book_Neely10} with $V=500$ to plot Figure \ref{fig:source-rate-convergence}.  It can be observed from Figure \ref{fig:source-rate-convergence} that Algorithm \ref{alg:new-alg} converges to the optimal source session rates faster than the backpressure algorithm in \cite{book_Neely10}.  The backpressure algorithm in \cite{book_Neely10} with $V=400$ takes around $2500$ iterations to converges to source rates close to $(1.2, 1.8)$ while  Algorithm \ref{alg:new-alg} only takes around $800$ iterations to converges to $(1.2, 1.8)$ (as shown in the zoom-in subfigure at the top right corner.)   In fact, the backpressure algorithm in \cite{book_Neely10} with $V=500$ can not converge to the exact optimal source session rate $(1.2, 1.8)$ but can only converge to its neighborhood with a distance gap determined by the value of $V$.  This is an effect from the fundamental $[O(1/V), O(V)]$ utility-delay tradeoff of the the backpressure algorithm in \cite{book_Neely10}. In contrast, Algorithm \ref{alg:new-alg} can eventually converge to the the exact optimal source session rate $(1.2, 1.8)$. A zoom-in subfigure at the bottom right corner in Figure \ref{alg:new-alg} verifies this and shows that the source rate for Session $1$ in Algorithm \ref{alg:new-alg} converges to $1.2$ while the source rate in the backpressure algorithm in \cite{book_Neely10} with $V=500$ oscillates around a point slightly larger than $1.2$.

\begin{figure}[htbp]
\centering
   \includegraphics[width=0.7\textwidth,height=0.7\textheight,keepaspectratio=true]{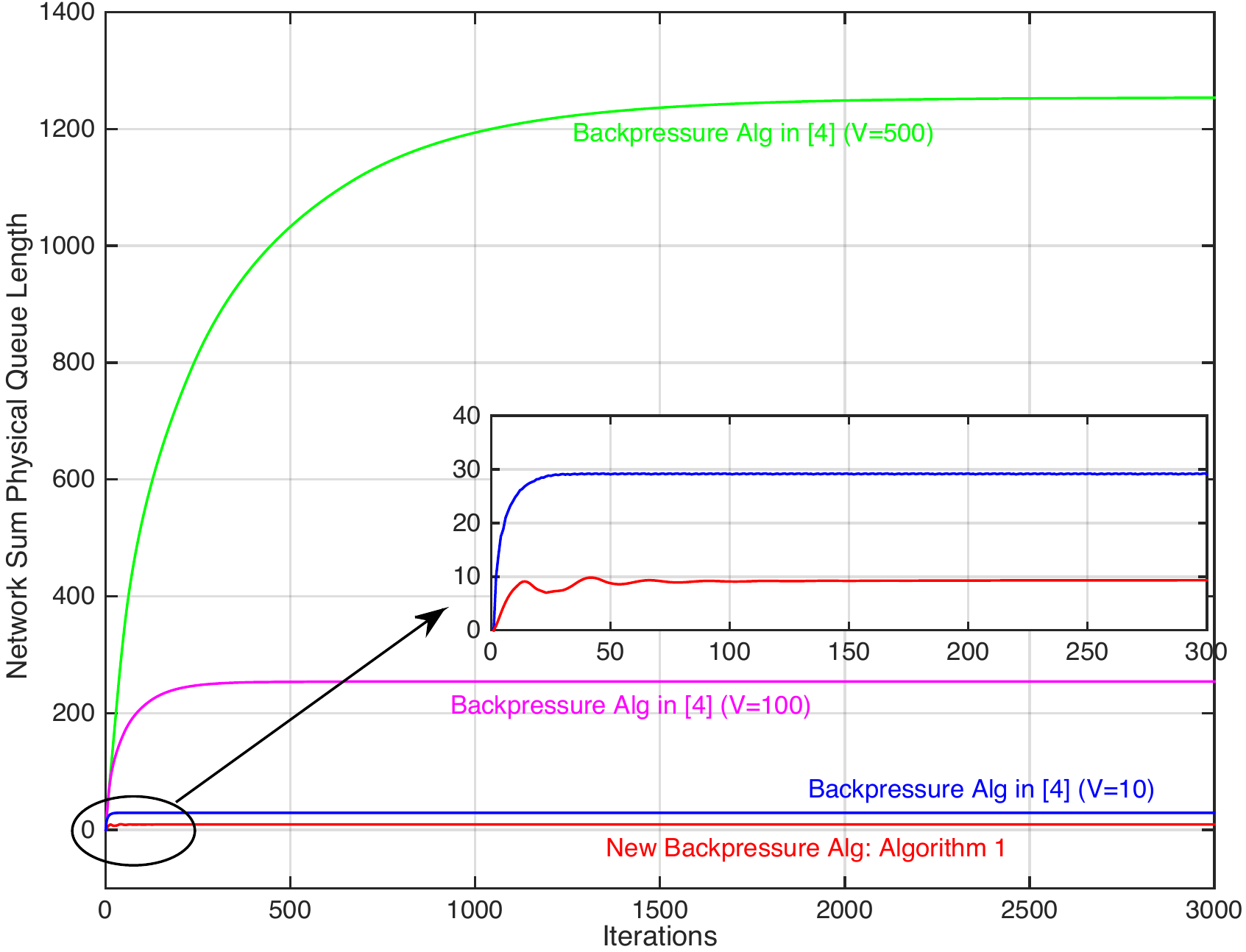} 
   \caption{Actual Queue length comparison between Algorithm \ref{alg:new-alg} and the backpressure algorithm in \cite{book_Neely10}.}
   \label{fig:queue}
\end{figure}

Corollary \ref{cor:physical-queue-bound} shows that Algorithm \ref{alg:new-alg} guarantees each actual queue in the network is bounded by constant $4\Vert \boldsymbol{\lambda}^{\ast}\Vert + 2\sqrt{ 2\xi} \Vert\mathbf{y}^{\ast}\Vert + \sum_{l\in \mathcal{O}(n)}C_{l}$. Recall that the backpressure algorithm in \cite{book_Neely10} can guarantee the actual queues in the network are bounded by a constant of order $V\Vert \boldsymbol{\lambda}^{\ast}\Vert $.  Figure \ref{fig:queue} plots the  sum of {\bf actual queue length} at each node for Algorithm \ref{alg:new-alg} and the backpressure algorithm in \cite{book_Neely10} with $V=10, 100$ and $500$. (Recall a larger $V$ in the backpressure algorithm in \cite{book_Neely10} yields a smaller utility gap but a larger queue length.)  It can be observed  that Algorithm \ref{alg:new-alg} has the smallest {\bf actual queue length} (see the zoom-in subfigure) and the actual queue length of the backpressure algorithm in \cite{book_Neely10} scales linearly with respect to $V$.

\section{Conclusion}
This paper develops a new first-order Lagrangian dual type backpressure algorithm for joint rate control and routing in multi-hop data networks.  The new backpressure algorithm can achieve vanishing utility optimality gaps and finite queue lengths. This improves the state-of-art $[O(1/V), O(V^2)]$ or $[O(1/V), O(V)]$ utility-delay tradeoff attained by existing backpressure algorithms \cite{Eryilmaz06JSAC,Neely05JSAC,LinShroff04CDC, LiuShroff15TON}.

\appendices

\section{Network Utility Maximization with Predetermined Multi-Path} \label{sec:multipath-num}
Consider multi-path network utility maximization in \cite{LinShroff06TAC} where each session has multiple given paths, then the source session rate $x_{f}$ in problem \eqref{eq:opt-obj}-\eqref{eq:opt-rate-nonnegative} becomes a vector $\mathbf{x}_{f} = [x_{f,j}]_{j\in \mathcal{P}_{f}}$ where $\mathcal{P}_{f}$ is the set of paths for session $f$ and the link session rate $\mu_{l}^{(f)}$ becomes a vector $\boldsymbol{\mu}_{l}^{(f)} = [\mu_{l}^{(f,j)}]_{j\in \mathcal{P}_{l}}$.  Define $\mathcal{S}_{l}^{(f)}$ as the set of paths for session $f$ that are allowed to use link $l$. Note that if all paths of session $f$ are forbidden to use link $l$, then $\mathcal{S}_{l}^{(f)}=\emptyset$. The multi-path network utility maximization problem can be formulated as follows:
\begin{align*}
\max&\sum_{f\in \mathcal{F}} U_{f}( \sum_{j\in \mathcal{P}_{f}}x_{f,j}) \\
\text{s.t.}~~  &   \sum_{j\in \mathcal{P}_{l}} x_{f,j} \mathbf{1}_{\{n=\text{Src}(f)\}} +  \sum_{j\in\mathcal{P}_{l}}\sum_{l\in \mathcal{I}(n)} \mu_{l}^{(f)} \leq    \sum_{j\in \mathcal{P}_{l}}\sum_{l\in \mathcal{O}(n)} \mu_{l}^{(f)}, \forall f\in \mathcal{F}, \forall n\in\mathcal{N}\setminus\{\text{Dst}(f)\}\\
			 &  \sum_{f\in \mathcal{F}} \sum_{j\in \mathcal{P}_{l}}\mu_{l}^{(f,j)}  \leq C_l, \forall l\in\mathcal{L}, \\
			 &\mu_{l}^{(f,j)} \geq 0, \forall l\in \mathcal{L},\forall f\in  \mathcal{F}, \forall j\in \mathcal{S}_{l}^{(f)},\\
			 &\mu_{l}^{(f,j)} = 0, \forall l\in \mathcal{L},\forall f\in \mathcal{F}, \forall j\in \mathcal{P}_{f} \setminus\mathcal{S}_{l}^{(f)}, \\
			 & \sum_{j\in \mathcal{P}_{l}}x_{f,j}\in \dom(U_f), \forall f\in \mathcal{F},\\
			 & x_{f,j} \geq 0, \forall f\in \mathcal{F}, j\in \mathcal{P}_f \end{align*}
The above formulation is in the form of problem  \eqref{eq:opt-obj}-\eqref{eq:opt-rate-nonnegative} except that the variable dimension is extended.

\section{An Example Illustrating the Possibly Large Gap between Model \eqref{eq:queue-update-model1} and Model \eqref{eq:queue-update-model2}} \label{sec:example-queue}

\begin{figure}[htbp]
\centering
   \includegraphics[width=0.5\textwidth,height=0.5\textheight,keepaspectratio=true]{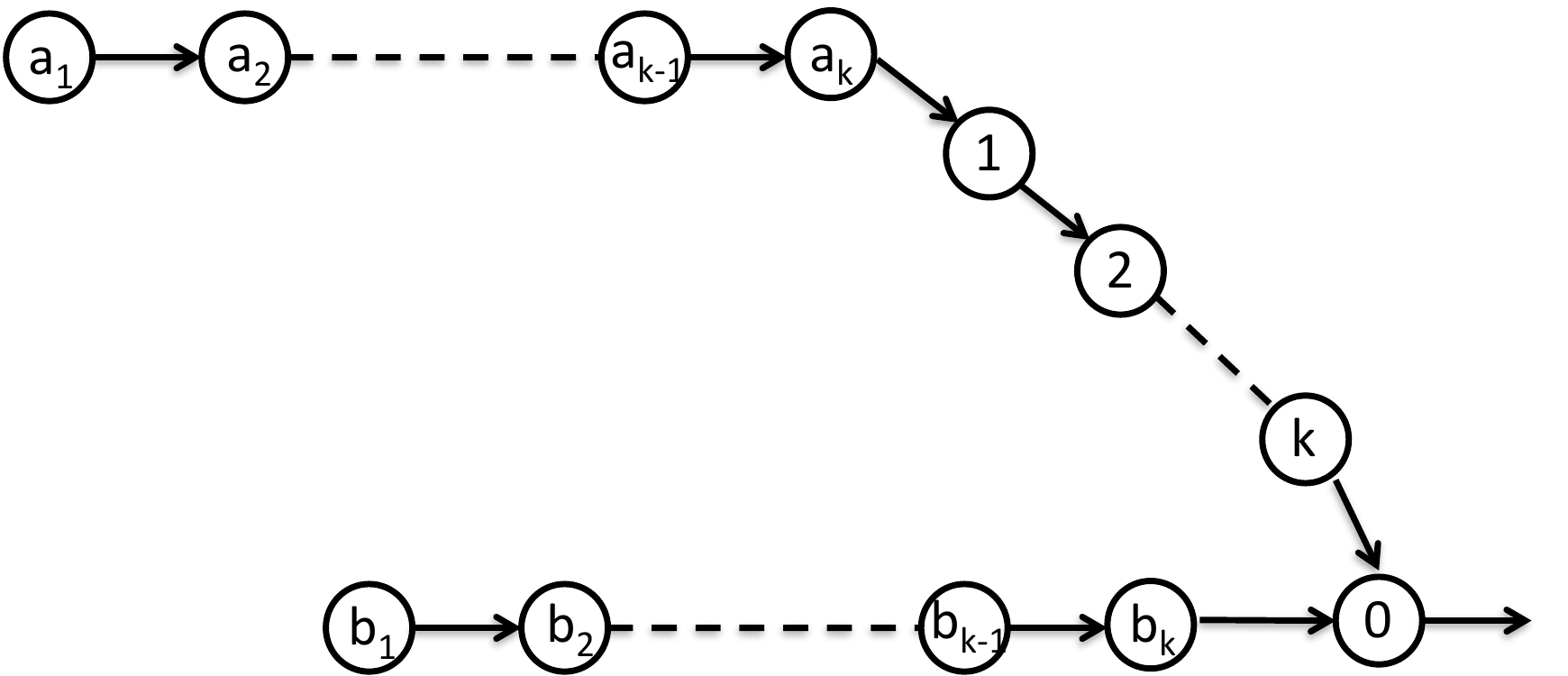} 
   \caption{An example illustrating the possibly large gap between queue model \eqref{eq:queue-update-model1} and  queue model \eqref{eq:queue-update-model2}}
   \label{fig:queue-model-exmple}
\end{figure}

Consider a network example shown in Figure \ref{fig:queue-model-exmple}. The network has $3k + 1$ nodes where only node $0$ is a destination; and $a_i, i\in \{1,2,\ldots, k\}$ and $b_i, i\in\{1,2,\ldots, k\}$ can have exogenous arrivals.  Assume all link capacities are equal to 1; and the exogenous arrivals are periodic with period 2k, as follows:
\begin{itemize}
\item Time slot $1$: One packet arrives at node $a_1$.
\item Time slot $2$: One packet arrives at node $a_2$.
\item $\cdots$
\item Time slot $k$: One packet arrives at node $a_k$.
\item Time slot $k + 1$: One packet arrives at node $b_1$.
\item Time slot $k + 2$: One packet arrives at node $b_2$.
\item $\cdots$
\item Time slot $2k$: One packet arrives at node $b_k$.
\end{itemize}

Under dynamics \eqref{eq:queue-update-model1}, each packet arrives on its own slot and traverses all links of its path to exit on the same slot it arrived. The queue backlog in each node is $0$ for all time.

Under dynamics \eqref{eq:queue-update-model2}, the first packet arrives at time slot $1$ to node $a_1$. This packet visits node $a_2$ at time slot $2$, when the second packet also arrives at $a_2$. One of these packets is delivered to node $a_3$ at time slot $3$, and another packet also arrives to node $3$. The nodes $\{1,\ldots, k\}$ do not have any exogenous arrivals and act only to delay the delivery of all packets from the ai nodes. It follows that the link from node $k$ to node $0$ will send exactly one packet over each slots $t\in\{2k+1,2k+2,\ldots,2k+k\}$. Similarly, the link from $b_k$ to $0$ sends exactly one packet to node $0$ over each of these same slots. Thus, node $0$ receives $2$ packets on each slot $t\in\{2k + 1,2k + 2, \ldots,2k + k\}$, but can only output $1$ packet per slot. The queue backlog in this node grows linearly and reaches $k + 1$ at time $2k + k$. Thus, the backlog in node $0$ can be arbitrarily large when $k$ is large. This example demonstrates that, even when there is only one destination, the deviation between virtual queues under dynamics \eqref{eq:queue-update-model1} and actual queues under dynamics \eqref{eq:queue-update-model2} can be arbitrarily large, even when the in-degree and out-degree of $1$ and an in-degree of at most $2$.

\section{Proof of Part (2) in Lemma \ref{lm:queue-model-relation}} \label{sec:proof-part2-lemma-ueue-model-relation}

Fix $f\in \mathcal{F}, n\in \mathcal{N}\setminus\{\text{Dst}(f)\}$.  By \eqref{eq:pf-queue-model-relation-eq1}, 
\begin{align*}
&\widehat{Q}_{n}^{(f)}[t+1] \\
=&\max\big\{ \widehat{Q}_{n}^{(f)}[t]  - \sum_{l\in \mathcal{O}(n)} \mu_{l}^{(f)}[t], 0\big\}+x_{f}[t] \mathbf{1}_{\{n=\text{Src}(f)\}}+ \sum_{l\in \mathcal{I}(n)} \mu_{l}^{(f)}[t] \\
=&\max\Big\{ \widehat{Q}_{n}^{(f)}[t]  +x_{f}[t] \mathbf{1}_{\{n=\text{Src}(f)\}} + \sum_{l\in \mathcal{I}(n)} \mu_{l}^{(f)}[t]  -\sum_{l\in \mathcal{O}(n)} \mu_{l}^{(f)}[t], ~~ x_{f}[t] \mathbf{1}_{\{n=\text{Src}(f)\}} + \sum_{l\in \mathcal{I}(n)} \mu_{l}^{(f)}[t]\Big\}\\
\overset{(a)}{\geq} &\max\Big\{ \widehat{Q}_{n}^{(f)}[t] +x_{f}[t] \mathbf{1}_{\{n=\text{Src}(f)\}} + \sum_{l\in \mathcal{I}(n)} \mu_{l}^{(f)}[t]  -\sum_{l\in \mathcal{O}(n)} \mu_{l}^{(f)}[t], ~~ 0\Big\}
\end{align*}
where (a) follows from the fact that $\mu_{l}^{(f)}[t], x_{f}[t], \forall f, l, t$ are non-negative. Note that the right side of the above equation is identical to the right side of \eqref{eq:queue-update-model1} and recall that $ Y_{n}^{(f)}[0] = 0 < \widehat{Q}_{n}^{(f)}[0]$. By inductions, we have  $Y_{n}^{(f)}[t] \leq \widehat{Q}_{n}^{(f)}[t], \forall t$. Since $\widehat{Q}_{n}^{(f)}[t] = Q_{n}^{(f)}[t] + B + \sum_{l\in \mathcal{O}(n)}C_{l}, \forall t$ and $Q_{n}^{(f)}[t]\leq B, \forall t$, we have $\widehat{Q}_{n}^{(f)}[t] \leq 2B + \sum_{l\in \mathcal{O}(n)}C_{l}, \forall t$. It follows that $Y_{n}^{(f)}[t] \leq 2B + \sum_{l\in \mathcal{O}(n)}C_{l}, \forall t$.

\section{Proof of Lemma \ref{lm:solution-routing-opt}}\label{sec:proof-solution-routing-opt}
Note that problem \eqref{eq:simple-routing-opt-obj}-\eqref{eq:simple-routing-opt-cons} satisfies Slater's condition. So the optimal solution to problem \eqref{eq:simple-routing-opt-obj}-\eqref{eq:simple-routing-opt-cons} is characterized by KKT conditions \cite{book_ConvexOptimization}.
Introducing Lagrange multipliers $\theta \in \mathbb{R}_{+}$ for inequality constraint $\sum_{k=1}^{K} z_k \leq b$ and $\boldsymbol{\nu} = [\nu_1, \ldots, \nu_{K}]\tran\in \mathbb{R}_+^{K}$ for inequality constraints $z_k\geq 0, k\in\{1,2,\ldots,K\}$. Let $\mathbf{z}^\ast = [z_{1}^{\ast}, \ldots, z_{K}^{\ast}]\tran$ and $(\theta^\ast, \boldsymbol{\nu}^\ast)$ be any primal and dual pair with the zero duality gap. By KKT conditions, we have $z_k^\ast - a_k + \theta^{\ast} - \nu_k^\ast = 0,\forall k\in\{1,2,\ldots, K\}; 
\sum_{k=1}^{K} z_{k}^\ast \leq  b;
\theta^{\ast} \geq 0;
\theta^{\ast} \big(\sum_{k=1}^{K} z_{k}^\ast -  b\big) = 0;
z_k^\ast \geq 0, \forall k\in\{1,2,\ldots, K\};
\nu_k^\ast \geq 0,  \forall k\in\{1,2,\ldots, K\};
\nu_k^\ast z_k^\ast =0,  \forall k\in\{1,2,\ldots, K\}$.

Eliminating $\nu_k^\ast, \forall k\in\{1,2,\ldots, K\}$ in all equations yields
$\theta^{\ast} \geq a_k - z_k^\ast, k\in\{1,2,\ldots, K\};
\sum_{k=1}^{K} z_{k}^\ast \leq  b;
\theta^{\ast} \geq 0;
\theta^{\ast} \big(\sum_{k=1}^{K} z_{k}^\ast -  b\big) = 0;
z_k^\ast \geq 0, \forall k\in\{1,2,\ldots, K\};
(z_k^\ast - a_k + \theta^{\ast}) z_k^\ast =0, \forall k\in\{1,2,\ldots, K\}$.

For all $ k\in\{1,2,\ldots, K\}$, we consider $\theta^{\ast} < a_k$ and $\theta^{\ast} \geq a_k$ separately:
\begin{enumerate}
\item If $\theta^{\ast} < a_k$ , then $\theta^{\ast} \geq a_k - z_k^\ast$ holds only when $z_k^\ast >0$, which by $(z_k^\ast - a_k + \theta^{\ast}) z_k^\ast =0$ implies that $z_k^\ast = a_k - \theta^{\ast}$.
\item If $\theta^{\ast} \geq a_k$, then $z_k^\ast >0$ is impossible, because $z_k^\ast >0$ implies that $z_k^\ast -a_k +\theta^{\ast} >0$, which together with $z_k^\ast >0$ contradicts the slackness condition $(z_k^\ast - a_k + \theta^{\ast}) z_k^\ast =0$. Thus, if $\theta^{\ast} \geq a_k$, we must have $z_k^\ast = 0$.
\end{enumerate}
Summarizing both cases, we have $z_k^\ast = \max\{0,a_k - \theta^{\ast}\}, \forall k\in\{1,2,\ldots,K\}$, where $\theta^{\ast}$ is chosen such that $\sum_{k=1}^{K} z_k^\ast \leq  b$, $\theta^{\ast} \geq 0$ and $\theta^{\ast} \big(\sum_{k=1}^{K} z_k^\ast -  b\big)= 0$.

To find such $\theta^{\ast}$, we first check if $\theta^{\ast} =0$. If $\theta^{\ast} =0$ is true, the slackness condition $\theta^{\ast} \big(\sum_{k=1}^{K} z_k^\ast -  b\big)$ is guaranteed to hold and we need to further require $\sum_{k=1}^{K} z_k^{\ast} = \sum_{k=1}^{K} \max\{0,a_{k}\}\leq b$. Thus $\theta^{\ast} =0$ if and only if $\sum_{k=1}^{K} \max \{0,a_{k}\} \leq b$. Thus, Algorithm \ref{alg:routing-update} check if $\sum_{k=1}^{K} \max\{0,a_{k}\} \leq b$ holds at the first step and if this is true, then we conclude $\theta^{\ast} = 0$ and we are done!

Otherwise, we know $\theta^{\ast} > 0$. By the slackness condition $\theta^{\ast} \big(\sum_{k=1}^{K} z_k^\ast -  b\big)=0$, we must have $\sum_{k=1}^{K} z_k^\ast = \sum_{k=1}^{K} \max\{0,a_{k} - \theta^{\ast}\} =b$. To find $\theta^{\ast}>0$ such that  $\sum_{k=1}^{K} \max\{0,a_{k} - \theta^{\ast}\} =b$, we could apply a bisection search by noting that all $z_k^{\ast}$ are decreasing with respect to $\theta^{\ast}$.

Another algorithm of finding $\theta^{\ast}$ is inspired by the observation that if $a_{j} \geq a_{i}, \forall i,j\in\{1,2,\ldots, K\}$, then $z_{j}^{\ast} \geq z_{i}^{\ast}$. Thus, we first sort all $a_{k}$ in a decreasing order, say $\pi$ is the permutation such that $a_{\pi(1)} \geq a_{\pi(2)} \geq \cdots \geq a_{\pi(K)}$; and then sequentially check if $k\in\{1,2,\ldots,K\}$ is the index such that $a_{\pi(k)} - \theta^{\ast} \geq 0$ and $a_{\pi(k+1)} - \theta^{\ast} < 0$. To check this, we first assume $k$ is indeed such an index and solve the equation $\sum_{j=1}^{k} ( a_{\pi(j)} - \theta^{\ast} ) = b$ to obtain $\theta^{\ast}$; (Note that in Algorithm \ref{alg:routing-update}, to avoid recalculating the partial sum $\sum_{j=1}^{k} a_{\pi(j)}$ for each $k$, we introduce the parameter $S_k =\sum_{j=1}^{k} a_{\pi(j)}$ and update $S_{k}$ incrementally. By doing this, the complexity of each iteration in the loop is only $O(1)$.) then verify the assumption by checking if $\theta^{\ast}\geq 0$, $a_{\pi(k)} - \theta^{\ast} \geq 0$ and $a_{\pi(k+1)} - \theta^{\ast} \leq 0$. The algorithm is described in Algorithm \ref{alg:routing-update} and has complexity $O(K\log(K))$. The overall complexity is dominated by the step of sorting all $a_{k}$.

\section{Proof of Lemma \ref{lm:alg-primal-opt}} \label{sec:proof-lemma-alg-primal-opt}
The objective function \eqref{eq:alg-opt-obj} can be rewritten as 
\begin{align}
&f(\mathbf{y}) -\sum_{\begin{subarray}{c}f\in \mathcal{F},\\ n\in \mathcal{N}\setminus \text{Dst}(f)\end{subarray}} \big(W_{n}^{(f)}[t] g_{n}^{(f)}(\mathbf{y}_{n}^{(f)}) + \alpha_{n} \Vert \mathbf{y}_{n}^{(f)} - \mathbf{y}_{n}^{(f)}[t-1]\Vert^{2} \big)-\sum_{f\in \mathcal{F}, n= \text{Dst}(f)}\alpha_{n} \sum_{l\in \mathcal{I}(n)} ( \mu_l^{(f)} -\mu_{l}^{(f)}[t-1])^{2} \nonumber\\
\overset{(a)}{=}&\sum_{f\in \mathcal{F}} U_{f}(x_{f}) - \sum_{\begin{subarray}{c}f\in \mathcal{F},\\ n\in \mathcal{N}\setminus \text{Dst}(f)\end{subarray}} W_{n}^{(f)}[t] \Big(x_{f} \mathbf{1}_{\{n=\text{Src}(f)\}}+ \sum_{l\in \mathcal{I}(n)} \mu_{l}^{(f)}- \sum_{l\in \mathcal{O}(n)} \mu_{l}^{(f)}\Big) \nonumber\\ & - \sum_{\begin{subarray}{c}f\in \mathcal{F},\\ n\in \mathcal{N}\setminus \text{Dst}(f)\end{subarray}} \alpha_{n} \Big((x_{f} - x_{f}[t-1])^{2} \mathbf{1}_{\{n=\text{Src}(f)\}} + \sum_{l\in \mathcal{I}(n)} (\mu_{l}^{(f)} -\mu_{l}^{(f)}[t-1])^{2} + \sum_{l\in\mathcal{O}(n)} (\mu_{l}^{(f)} - \mu_{l}^{(f)}[t-1])^{2}\Big) \nonumber\\&-\sum_{f\in \mathcal{F}, n= \text{Dst}(f)}\alpha_{n} \sum_{l\in \mathcal{I}(n)} ( \mu_l^{(f)} -\mu_{l}^{(f)}[t-1])^{2} \nonumber\\
\overset{(b)}{=} &\sum_{f\in \mathcal{F}} \big(U_{f}(x_{f})  - W_{\text{Src}(f)}^{(f)} [t]x_{f} - \alpha_{\text{Src}(f)} (x_{f} - x_{f}[t-1])^{2}\big) + \sum_{(n,m)\in \mathcal{L}}  \sum_{f\in \mathcal{F}} \big( W_{n}^{(f)}[t] - W_{m}^{(f)}[t] \big) \mu_{(n,m)}^{(f)}  \nonumber\\&-  \sum_{(n,m)\in \mathcal{L}} (\alpha_{n} + \alpha_{m})\sum_{f\in \mathcal{F}} (\mu_{(n,m)}^{(f)} - \mu_{(n,m)}^{(f)}[t-1])^{2}
\label{eq:pf-lemma-alg-primal-opt}
\end{align}
where (a) follows from the fact that $g_{n}^{(f)}(\mathbf{y}_{n}^{(f)}) = x_{f} \mathbf{1}_{\{n=\text{Src}(f)\}}+ \sum_{l\in \mathcal{I}(n)} \mu_{l}^{(f)}- \sum_{l\in \mathcal{O}(n)} \mu_{l}^{(f)}$ and $\Vert \mathbf{y}_{n}^{(f)} - \mathbf{y}_{n}^{(f)}[t-1]\Vert^{2} = (x_{f} - x_{f}[t-1])^{2} \mathbf{1}_{\{n=\text{Src}(f)\}} + \sum_{l\in \mathcal{I}(n)} (\mu_{l}^{(f)} -\mu_{l}^{(f)}[t-1])^{2} + \sum_{l\in\mathcal{O}(n)} (\mu_{l}^{(f)} - \mu_{l}^{(f)}[t-1])^{2}$; and (b) follows by collecting each linear term $\mu_{l}^{(f)}$ and each quadratic term $(\mu_{l}^{(f)} - \mu_{l}^{(f)}[t-1])^{2}$. Note that each link session rate $\mu_{l}^{(f)}$ appears twice with opposite signs in the summation term $\sum_{f\in \mathcal{F},n\in \mathcal{N}\setminus\{\text{Dst}(f)\}} W_{n}^{(f)}[t] \big(x_{f} \mathbf{1}_{\{n=\text{Src}(f)\}} + \sum_{l\in \mathcal{I}(n)} \mu_{l}^{(f)}- \sum_{l\in \mathcal{O}(n)} \mu_{l}^{(f)}\big)$ unless link $l$ flows into $\text{Dst}(f)$ and recall that $W_{\text{Dst}(f)}^{(f)} = 0, \forall f\in \mathcal{F}$. The quadratic terms are collected in a similar way. Note that the term $\sum_{f\in \mathcal{F}, n= \text{Dst}(f)}\alpha_{n} \sum_{l\in \mathcal{I}(n)} ( \mu_l^{(f)} -\mu_{l}^{(f)}[t-1])^{2}$ introduced to the objective function \eqref{eq:alg-opt-obj} is necessary to guarantee each quadratic term $(\mu_{(m,n)}^{(f)} - \mu_{(m,n)}^{(f)}[t-1])^{2}$ with the same link index $(n,m)$ but different flow indices $f\in \mathcal{F}$ have the same coefficient $\alpha_n +\alpha_m$ in the last line of \eqref{eq:pf-lemma-alg-primal-opt}. 

Note that equation \eqref{eq:pf-lemma-alg-primal-opt} is now separable for each scalar $x_{f}$ and vector $[\mu_{(n,m)}^{(f)}]_{f\in \mathcal{F}}$. Thus, problem \eqref{eq:alg-opt-obj}-\eqref{eq:alg-opt-cons} can be decomposed into independent smaller optimization problems in the form of problem \eqref{eq:source-opt-obj}-\eqref{eq:souce-opt-cons} with respect to each scalar $x_{f}$, and in the form of problem \eqref{eq:routing-opt-obj}-\eqref{eq:routing-opt-cons-zero} with respect to each vector $[\mu_{(n,m)}^{(f)}]_{f\in \mathcal{F}}$.

\section{Proof of Lemma \ref{lm:dpp-bound}} \label{sec:proof-lemma-dpp-bond}
Note that $W_{n}^{(f)}[t]$ appears as a known constant in \eqref{eq:source-opt-obj}. Since $U_{f}(x_{f})$ is concave and $W_{n}^{(f)}[t] x_{f} $ is linear, it follows that \eqref{eq:source-opt-obj} is strongly concave with respect to $x_{f}$ with modulus $2\alpha_{n}$. Since $x_{f}[t]$ is chosen to solve  \eqref{eq:source-opt-obj}-\eqref{eq:souce-opt-cons}, by Lemma \ref{lm:strong-convex-quadratic-optimality}, $\forall f\in \mathcal{F}$, we have 
\begin{align}
&\underbrace{U_{f}(x_{f}[t]) - W_{\text{Src}(f)}^{(f)}[t] x_{f}[t] - \alpha_{n} (x_{f}[t] - x_{f}[t-1])^{2}}_{\text{\eqref{eq:pf-dpp-bound-eq1}-I}} 
\geq \underbrace{U_{f}(x_{f}^{\ast}) - W_{\text{Src}(f)}^{(f)}[t] x_{f}^{\ast} - \alpha_{n} (x_{f}^{\ast} - x_{f}[t-1])^{2} + \alpha_{n} (x_{f}^{\ast} - x_{f}[t])^{2}}_{\text{\eqref{eq:pf-dpp-bound-eq1}-II}}. \label{eq:pf-dpp-bound-eq1}
\end{align}

Similarly, we know \eqref{eq:routing-opt-obj} is strongly concave with respect to vector $[\mu_{(n,m)}^{f}]_{f\in \mathcal{F}}$ with modulus $2(\alpha_{n} + \alpha_{m})$. By Lemma \ref{lm:strong-convex-quadratic-optimality}, $\forall (n,m)\in \mathcal{O}(n)$, we have
\begin{align}
&\underbrace{\sum_{f\in \mathcal{F}}  \big( W_{n}^{(f)}[t]  - W_{m}^{(f)}[t]\big) \mu_{(n,m)}^{(f)}[t] -\big(\alpha_{n} + \alpha_{m}\big)\sum_{f\in\mathcal{F}} \big(\mu_{(n,m)}^{(f)}[t] -\mu_{(n,m)}^{(f)}[t-1]\big)^{2}}_{\text{\eqref{eq:pf-dpp-bound-eq2}-I}} \nonumber\\
\geq &\underbrace{\sum_{f\in \mathcal{F}}  \big( W_{n}^{(f)}[t]  - W_{m}^{(f)}[t]\big) \mu_{(n,m)}^{(f),\ast} -\big(\alpha_{n} + \alpha_{m}\big)\sum_{f\in\mathcal{F}} \big(\mu_{(n,m)}^{(f),\ast} -\mu_{(n,m)}^{(f)}[t-1]\big)^{2} + \big(\alpha_{n} + \alpha_{m}\big)\sum_{f\in\mathcal{F}} \big(\mu_{(n,m)}^{(f),\ast} -\mu_{(n,m)}^{(f)}[t]\big)^{2}}_{\text{\eqref{eq:pf-dpp-bound-eq2}-II}}. \label{eq:pf-dpp-bound-eq2}
\end{align}

Recall that each column vector $\mathbf{y}_{n}^{(f)} $ defined in \eqref{eq:y-each-flow-balance-cons} is composed by control actions that appear in each constraint \eqref{eq:opt-flow-balance-cons}; column vector $\mathbf{y}=[x_{f}; \mu_{l}^{(f)}]_{f\in \mathcal{F}, l\in \mathcal{L}}$ is the collection of all control actions; and $f(\mathbf{y}) = \sum_{f\in \mathcal{F}} U_{f}(x_{f})$. Summing term \eqref{eq:pf-dpp-bound-eq1}-I over all $f\in \mathcal{F}$ and term \eqref{eq:pf-dpp-bound-eq2}-I over all $(n,m)\in \mathcal{L}$ and using an argument similar to the proof of Lemma \ref{lm:alg-primal-opt} (Recall that $\mathbf{y}[t]$ is jointly chosen to minimize \eqref{eq:alg-opt-obj} by Lemma \ref{lm:alg-primal-opt}.) yields 
\begin{align}
&\sum_{f\in \mathcal{F}} \text{\eqref{eq:pf-dpp-bound-eq1}-I} + \sum_{(n,m)\in \mathcal{N}} \text{ \eqref{eq:pf-dpp-bound-eq2}-I}\nonumber\\
=& f(\mathbf{y}[t])-\sum_{\begin{subarray}{c}f\in \mathcal{F},\\ n\in \mathcal{N}\setminus \text{Dst}(f)\end{subarray}}  \big(W_{n}^{(f)}[t] g_{n}^{(f)}(\mathbf{y}_{n}^{(f)}[t]) + \alpha_{n}\Vert \mathbf{y}_{n}^{(f)}[t] - \mathbf{y}_{n}^{(f)}[t-1]\Vert^{2}\big) - \sum_{\begin{subarray}{c}f\in \mathcal{F},\\ n=\text{Dst}(f)\end{subarray}}\alpha_{n} \sum_{l\in \mathcal{I}(n)} ( \mu_l^{(f)}[t] -\mu_{l}^{(f)}[t-1])^{2}.\label{eq:pf-dpp-bound-eq3}
\end{align}

Recall that $\Phi[t] =\sum_{f\in \mathcal{F}, n\in \mathcal{N}} \big(\alpha_{n}\mathbf{1}_{\{n\neq \text{Dst}(f)\}} \Vert \mathbf{y}_{n}^{(f),\ast} - \mathbf{y}_{n}^{(f)}[t]\Vert^{2}  + \alpha_{n} \mathbf{1}_{\{n=\text{Dst}(f)\}}\sum_{l\in \mathcal{I}(n)}( \mu_l^{(f),\ast} -\mu_{l}^{(f)}[t])^{2} \big)$. Summing term \eqref{eq:pf-dpp-bound-eq1}-II over all $f\in \mathcal{F}$ and term \eqref{eq:pf-dpp-bound-eq2}-II over all $(n,m)\in \mathcal{L}$ yields
\begin{align}
\sum_{f\in \mathcal{F}} \text{\eqref{eq:pf-dpp-bound-eq1}-II} + \sum_{(n,m)\in \mathcal{N}} \text{ \eqref{eq:pf-dpp-bound-eq2}-II}
= f(\mathbf{y}^{\ast}) +\Phi[t] - \Phi[t-1]-\sum_{\begin{subarray}{c}f\in \mathcal{F},\\ n\in \mathcal{N}\setminus \text{Dst}(f)\end{subarray}} W_{n}^{(f)}[t] g_{n}^{(f)}(\mathbf{y}_{n}^{(f),\ast})  ,\label{eq:pf-dpp-bound-eq4}
\end{align}

Combining \eqref{eq:pf-dpp-bound-eq1}-\eqref{eq:pf-dpp-bound-eq4}  and rearranging terms yields 
\begin{align}
&f(\mathbf{y}[t]) \nonumber \\
\geq &  f(\mathbf{y}^{\ast}) +\Phi[t] - \Phi[t-1] -\sum_{\begin{subarray}{c}f\in \mathcal{F},\\ n\in \mathcal{N}\setminus \text{Dst}(f)\end{subarray}}  W_{n}^{(f)}[t] g_{n}^{(f)}(\mathbf{y}_{n}^{(f),\ast})   +\sum_{\begin{subarray}{c}f\in \mathcal{F},\\ n\in \mathcal{N}\setminus \text{Dst}(f)\end{subarray}}  \big(W_{n}^{(f)}[t] g_{n}^{(f)}(\mathbf{y}_{n}^{(f)}[t]) + \alpha_{n}\Vert \mathbf{y}_{n}^{(f)}[t] - \mathbf{y}_{n}^{(f)}[t-1]\Vert^{2}\big) \nonumber\\ &+ \sum_{\begin{subarray}{c}f\in \mathcal{F},\\ n= \text{Dst}(f)\end{subarray}}\alpha_{n} \sum_{l\in \mathcal{I}(n)} ( \mu_l^{(f)}[t] -\mu_{l}^{(f)}[t-1])^{2} \nonumber\\
\overset{(a)}{\geq} & f(\mathbf{y}^{\ast}) + \Phi[t] -\Phi[t-1] + \sum_{\begin{subarray}{c}f\in \mathcal{F},\\ n\in \mathcal{N}\setminus \text{Dst}(f)\end{subarray}}  \big(W_{n}^{(f)}[t] g_{n}^{(f)}(\mathbf{y}_{n}^{(f)}[t]) + \alpha_{n}\Vert \mathbf{y}_{n}^{(f)}[t] - \mathbf{y}_{n}^{(f)}[t-1]\Vert^{2}\big) \nonumber\\
\overset{(b)} {=}& f(\mathbf{y}^{\ast}) + \Phi[t] -\Phi[t-1] + \sum_{\begin{subarray}{c}f\in \mathcal{F},\\ n\in \mathcal{N}\setminus \text{Dst}(f)\end{subarray}}  \big(Q_{n}^{(f)}[t] g_{n}^{(f)}(\mathbf{y}_{n}^{(f)}[t]) +g_{n}^{(f)}(\mathbf{y}_{n}^{(f)}[t-1])g_{n}^{(f)}(\mathbf{y}_{n}^{(f)}[t])  + \alpha_{n}\Vert \mathbf{y}_{n}^{(f)}[t] - \mathbf{y}_{n}^{(f)}[t-1]\Vert^{2}\big)  \label{eq:pf-dpp-bound-eq5}
\end{align} 
where (a) follows because $g_{n}^{(f)}(\mathbf{y}_{n}^{(f),\ast}) = 0, \forall f\in \mathcal{F}, \forall n\in \mathcal{N}\setminus \text{Dst}(f),$ and $\sum_{f\in \mathcal{F},n=\text{Dst}(f)}\alpha_{n} \sum_{l\in \mathcal{I}(n)} ( \mu_l^{(f)}[t] -\mu_{l}^{(f)}[t-1])^{2}\geq 0$; (b) follows from the fact that $W_{n}^{(f)}[t] = Q_{n}^{(f)}[t] + g_{n}^{(f)}(\mathbf{y}_{n}^{(f)}[t-1])$.

Recall that $u_{1}\tran u_{2} = \frac{1}{2} u_{1}^{2}  + \frac{1}{2} u_{2}^{2} - \frac{1}{2} (u_{1} -u_{2})^{2}$ for any $u_{1}, u_{2}\in \mathbb{R}$. Thus, for all $f\in \mathcal{F}, n\in \mathcal{N}\setminus\text{Dst}(f)$, we have 
\begin{align}
g_{n}^{(f)}(\mathbf{y}_{n}^{(f)}[t-1]) g_{n}^{(f)}(\mathbf{y}_{n}^{(f)}[t]) 
 = \frac{1}{2} \big(g_{n}^{(f)}(\mathbf{y}_{n}^{(f)}[t-1]) \big)^{2}  + \frac{1}{2}\big(g_{n}^{(f)}(\mathbf{y}_{n}^{(f)}[t]) \big)^{2} - \frac{1}{2}\big(g_{n}^{(f)}(\mathbf{y}_{n}^{(f)}[t-1]) - g_{n}^{(f)}(\mathbf{y}_{n}^{(f)}[t])  \big)^{2} . \label{eq:pf-dpp-bound-eq6}
\end{align} 
Substituting \eqref{eq:pf-dpp-bound-eq6} into \eqref{eq:pf-dpp-bound-eq5} yields 
\begin{align}
f(\mathbf{y}[t])
\geq & f(\mathbf{y}^{\ast}) + \Phi[t] -\Phi[t-1] + \sum_{\begin{subarray}{c}f\in \mathcal{F},\\ n\in \mathcal{N}\setminus \text{Dst}(f)\end{subarray}}  \Big( Q_{n}^{(f)}[t] g_{n}^{(f)}(\mathbf{y}_{n}^{(f)}[t]) +\frac{1}{2} \big(g_{n}^{(f)}(\mathbf{y}_{n}^{(f)}[t-1]) \big)^{2}  + \frac{1}{2}\big(g_{n}^{(f)}(\mathbf{y}_{n}^{(f)}[t]) \big)^{2} \nonumber \\ &- \frac{1}{2}\big(g_{n}^{(f)}(\mathbf{y}_{n}^{(f)}[t-1]) - g_{n}^{(f)}(\mathbf{y}_{n}^{(f)}[t])  \big)^{2} + \alpha_{n}\Vert \mathbf{y}_{n}^{(f)}[t] - \mathbf{y}_{n}^{(f)}[t-1]\Vert^{2}\Big) \nonumber\\
\overset{(a)}{\geq } &f(\mathbf{y}^{\ast}) + \Phi[t] -\Phi[t-1] + \sum_{\begin{subarray}{c}f\in \mathcal{F},\\ n\in \mathcal{N}\setminus \text{Dst}(f)\end{subarray}}  \Big( Q_{n}^{(f)}[t] g_{n}^{(f)}(\mathbf{y}_{n}^{(f)}[t]) +\frac{1}{2} \big(g_{n}^{(f)}(\mathbf{y}_{n}^{(f)}[t-1]) \big)^{2}  + \frac{1}{2}\big(g_{n}^{(f)}(\mathbf{y}_{n}^{(f)}[t]) \big)^{2} \nonumber \\ & +\big(\alpha_{n} -\frac{1}{2}\beta_{n}^{2}\big) \Vert \mathbf{y}_{n}^{(f)}[t] - \mathbf{y}_{n}^{(f)}[t-1]\Vert^{2}\Big) \nonumber\\
\overset{(b)}{\geq } &f(\mathbf{y}^{\ast}) + \Phi[t] -\Phi[t-1] + \sum_{\begin{subarray}{c}f\in \mathcal{F},\\ n\in \mathcal{N}\setminus \text{Dst}(f)\end{subarray}}  \Big( Q_{n}^{(f)}[t] g_{n}^{(f)}(\mathbf{y}_{n}^{(f)}[t]) + \frac{1}{2}\big(g_{n}^{(f)}(\mathbf{y}_{n}^{(f)}[t]) \big)^{2}\Big)
\label{eq:pf-dpp-bound-eq7}
\end{align}
where (a) follows from the Fact \ref{fact:flow-cons-Lipschitz}, i.e., each $g_{n}^{(f)}(\cdot)$ is Lipschitz with modulus $\beta_{n}$ and (b) follows because $\alpha_{n}\geq \frac{1}{2}(d_{n}+1)$, $\beta_{n}\leq \sqrt{d_{n}+1}$ and $\frac{1}{2} \big(g_{n}^{(f)}(\mathbf{y}_{n}^{(f)}[t-1]) \big)^{2}\geq 0$.

Subtracting \eqref{eq:drift-equality} from \eqref{eq:pf-dpp-bound-eq7} and cancelling the common terms on both sides  yields
\begin{align*}
-\Delta[t] + f(\mathbf{y}[t]) \geq f(\mathbf{y}^{\ast}) + \Phi[t] - \Phi[t-1].
\end{align*}

\bibliographystyle{IEEEtran}
\bibliography{IEEEfull,mybibfile}

\end{document}